\documentclass{aa}
\usepackage{graphicx}
\usepackage[varg]{txfonts}
\usepackage{natbib}
\usepackage{subfig}
\usepackage{longtable}
\newcommand{\bd}{BD+46$^{\circ}$442}
\newcommand{\angstrom}{\AA}
\newcommand{\teff}{T_\mathrm{eff}}
\newcommand{\dd}{\mathrm{d}}

\begin{document}
   \title{ Jet creation in post-AGB binaries: the circum-companion accretion disc around \bd.
\thanks{Based on observations made with the Mercator Telescope, operated on the island of La Palma by the Flemmish Community, at the Spanish Observatorio del Roque de los Muchachos of the Instituto de Astrofísica de Canarias.}
\thanks{Reduced spectra are only available in electronic form at the CDS via anonymous ftp to cdsarc.u-strasbg.fr (130.79.128.5) or via http://cdsweb.u-strasbg.fr/cgi-bin/qcat?J/A+A/}}
   \titlerunning{Jet formation in \bd}
   \author{Dylan Bollen \inst{1} 
           \and 
           Hans Van Winckel \inst{1}
           \and
           Devika Kamath \inst{1,2,3}
         }
   \institute{Instituut voor Sterrenkunde (IvS), KU Leuven,
              Celestijnenlaan 200D, B-3001 Leuven, Belgium\\
              \email{dylan.bollen@kuleuven.be}
              \and
              Department of Physics \& Astronomy, Macquarie University, 
              Sydney, NSW 2109, Australia
              \and
              Astronomy, Astrophysics and Astrophotonics Research Centre, 
              Macquarie University, Sydney, NSW 2109, Australia
              }
   
   \date{Received: 3 July, 2017 / Accepted: 28 July, 2017}
   \authorrunning{Bollen et al.}

  \abstract
  {}
  {We aim at describing and understanding binary interaction processes in systems with very evolved companions. Here, we focus on understanding the origin and determining the properties of the high-velocity outflow observed in one such system.}
  {We present a quantitative analysis of \bd, a post-AGB binary that shows active mass transfer that leads to the creation of a disk-driven outflow or jet. We obtained high-resolution optical spectra from the HERMES spectrograph, mounted on the $1.2\,$m Flemish \textit{Mercator} Telescope. By performing a time-series analysis of the H$\alpha$ profile, we identified the different components of the system. We deduced the jet geometry by comparing the orbital phased data with our jet model. In order to image the accretion disk around the companion of \bd, we applied the technique of Doppler tomography.}
  {The orbital phase-dependent variations in the H$\alpha$ profile can be related to an accretion disk around the companion, from which a high-velocity outflow or jet is launched. Our model shows that there is a clear correlation between the inclination angle and the jet opening angle. The latitudinally dependent velocity structure of our jet model shows a good correspondence to the data, with outflow velocities higher than at least 400$\,$km$\,$s$^{-1}$. The intensity peak in the Doppler map might be partly caused by a hot spot in the disk, or by a larger asymmetrical structure in the disk.}
  {We show that \bd\, is a result of a binary interaction channel. The origin of the fast outflow in this system might be to a gaseous disk around the secondary component, which is most likely a main-sequence star. Our analysis suggests that the outflow has a rather wide opening angle and is not strongly collimated. Our time-resolved spectral monitoring reveals the launching site of the jet in the binary \bd. Similar orbital phase-dependent H$\alpha$ profiles are commonly observed in post-AGB binaries. Post-AGB binaries provide ideal test beds to study jet formation and launching mechanisms over a wide range of orbital conditions.}  

   \maketitle
%
%________________________________________________________________

\section{Introduction}
Planetary nebulae (PNe) and proto-planetary nebulae (PPNe) display a wide variety in shapes; they have point-symmetric, bipolar, or even multipolar structures. Jet formation is an important mechanism for the shaping of the circumstellar material in PNe, but it still remains a poorly understood phenomenon \citep{jones17a}. In this model, a collimated high-velocity outflow or jet operates during the late asymptotic giant branch (AGB) or early post-AGB phase and leads to the observed asymmetric PNe and PPNe \citep{sahai98c}.

Two proposed mechanisms for the formation of these jets, which are not mutually exclusive, are the presence of magnetic fields \citep{garcia05} and a close binary companion \citep{morris87}. \citet{soker06} showed that a single star cannot supply the energy and angular momentum carried by the magnetic fields in order to shape the circumstellar outflows. Hence, a companion is needed. The companion in the binary system accretes matter from the evolved star either by accreting the AGB wind or through Roche-lobe overflow (RLOF) \citep[][and references therein]{demarco09}. This induces the formation of an accretion disk around the companion, facilitating a disk-driven outflow, where the presence of a magnetic field could also play an important role in the launching and collimation of the jets \citep{nordhaus07}.

\citet{blackman14} determined the minimum accretion rates for 19 PPNe from their observed outflow momenta, assuming a main-sequence (MS) star or white dwarf (WD) as accretor. These accretion rates were compared with theoretical accretion models, from which they concluded that Bondi-Hoyle-Lyttleton wind accretion and wind RLOF are too weak to establish the outflow. The mechanisms that are capable of powering the outflows of these objects are RLOF and accretion during the common envelope evolution.\\

The central regions of PNe and PPNe are highly embedded, however. Hence, the difficulty of probing the inner regions of these systems makes it complicated to study the origin and properties of the mechanisms responsible for their shaping. In order to acquire more knowledge in these fundamental questions, it is interesting to study optically bright post-AGB stars that also show the presence of a jet. The post-AGB is only a short-lived phase in the lifetime of the star, during which the central star will become hotter and is surrounded by the ejected circumstellar material. For optically bright post-AGB stars with a near-infrared excess in their spectral energy distribution, this material will most often reside in a circumstellar Keplerian disk \citep{vanwinckel03}. Radial velocity measurements of post-AGB stars with a disk have revealed a binary nature for several of these systems \citep[e.g.][]{waelkens91,waters93,vanwinckel95}. Since this discovery, long monitoring programmes have been initiated \citep[e.g.][]{vanwinckel09} in order to detect more systems that show binarity. By now, it is generally accepted that all post-AGB stars with a disk-like structure are part of a binary system.\\

Most post-AGB binaries are not spatially resolved and do not show a nebula. These objects are called naked post-AGB stars \citep[e.g.][and references therein]{lagadec17}. An exception is the Red Rectangle nebula. This is a bipolar post-AGB star, surrounding HD 44179 \citep{cohen75}, for which a binary was discovered by \citet{waelkens96} and which also shows the presence of jets. The system contains an X-shaped biconical outflow, which emerges from an optically thick disk \citep{menshchikov98, cohen04}. Since the disk is seen nearly edge-on, the central region is observed by scattering of the photons above and below the disk. \citet{witt09} presented the results of a seven-year spectroscopic program of the RR. From the variations in the H$\alpha$ line profile, they revealed that the secondary component in this system is the origin of the fast bipolar outflow, which reaches velocities of up to $560\,$km s$^{-1}$. The accretion disk surrounding the secondary is identified as the source of the Lyman/far-UV continuum.

The study of the RR shows that it is possible to obtain knowledge about the origin and properties of jets. In addition to the RR, the presence of a jet is described in another post-AGB binary as well, namely \bd\, \citep{gorlova12}. This is a naked post-AGB binary, surrounded by a circumbinary Keplerian disk. As for the RR, the H$\alpha$ line profile variations are related to a high-velocity outflow or jet. The main advantage of this system is that its central region is not highly obscured and can thus be observed directly, in contrast to the RR, which is seen in scattered light. These ideal observational conditions make systems such as \bd\, optimal to study the physical mechanisms that lead to jet creation.\\
%While the link to PNe and post-AGB binaries is not clear \citep{vanwinckel03}, these systems provide ideal observational conditions to study jet creation.\\

In this work, we present the results of our detailed modelling of the post-AGB binary \bd. The paper is organised as follows: we introduce the observations (Sect. \ref{sec:obs}) and the object (Sect. \ref{sec:prop}). We present our time-series analyses of the H$\alpha$-line and several other line profiles in Sect. \ref{sec:tsa}. A deduction of the geometry of the jet is presented in Sect. \ref{sec:jetgeom}. In Sect. \ref{doptom}, the technique of Doppler tomography is applied in order to resolve the circum-companion accretion disk. We discuss these results in Sect. \ref{sec:concl}.

\section{Properties of \bd}\label{sec:prop}
The source \bd\, is a G-type metal-poor evolved star. \citet{gorlova12} estimated stellar parameters from the spectra by comparing the hydrogen and metal lines with synthetic profiles. Their estimated stellar parameters are given in Table \ref{tab:BDstelpar}. Furthermore, they found that the spectral energy distribution (SED) of \bd\, shows a clear excess emission beyond $\lambda = 2\,\mu$m, with a downwards slope towards higher wavelengths \citep[see Fig. 6 in ][]{gorlova12}. This is also a clear indication of a post-AGB binary with a circumbinary disk \citep[e.g.][]{deruyter06, manick17}.
\begin{table}
\begin{center}   
  \caption{Effective temperature $\teff$, surface gravity $\log g$, and metallicity [M/H] of \bd\, derived by \citet{gorlova12}. }\label{tab:BDstelpar}
\begin{footnotesize}
\begin{tabular}{ccc} \hline \hline
$\teff$ & $\log g$ & [M/H]\\
   (K) &  & (dex)\\
\hline
$6250\pm 250$ & $1.5\pm 0.5$ & $-0.7\pm 0.2$\\
\hline
\end{tabular}
\end{footnotesize}
\end{center}
\end{table}
\citet{gorlova12} also determined the orbital parameters of \bd\, from the radial velocity (RV) variations of 60 spectra. They derived an orbital period of $P = 140.77\pm0.02\,$d, with an eccentricity of $e = 0.08\pm 0.002$ and an RV of $K_1 = 23.66\,$km s$^{-1}$ for the giant. The periodic RV variations in combination with the peak-to-peak difference in RV are a clear indication of a Keplerian orbit of the giant instead of pulsations. The companion has an estimated RV amplitude of $K_2 = 15\pm 5\,$km s$^{-1}$. This leads to a mass ratio of $q = K_1/K_2 = 0.6$ and indicates the companion to be the more massive component, with the giant being a post-AGB star that went through a significant period of mass loss. We assume this mass ratio for further calculations.\\

\begin{figure}[t!]
\centering
\includegraphics[width=.5\textwidth]{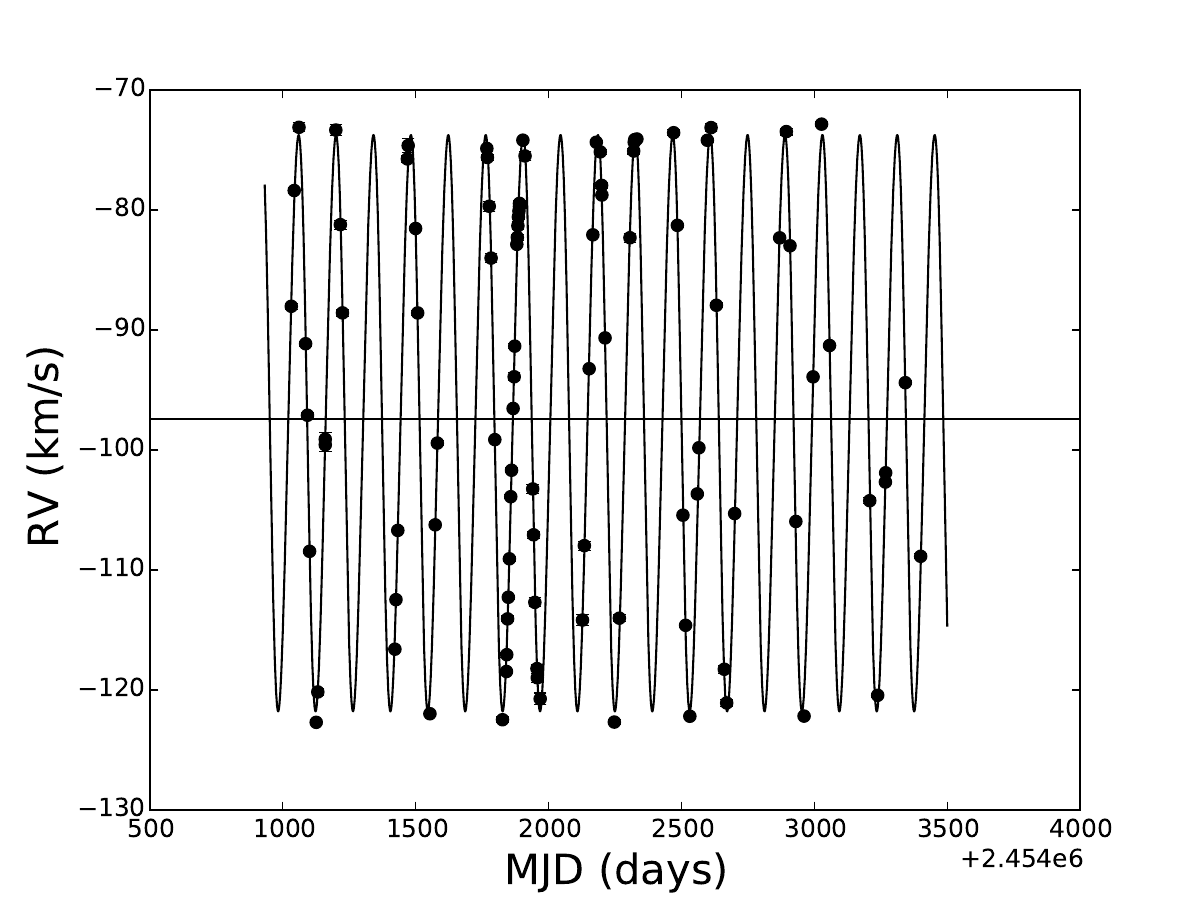}
\caption{Radial velocities of the primary component of BD+46$\degr$442. The full line denotes the RV curve for a Keplerian orbit with the orbital parameters given in Table~\ref{tab:bdorbpar}}\label{fig:rvtime}
\end{figure}

\begin{figure}[t!]
\centering
\includegraphics[width=.5\textwidth]{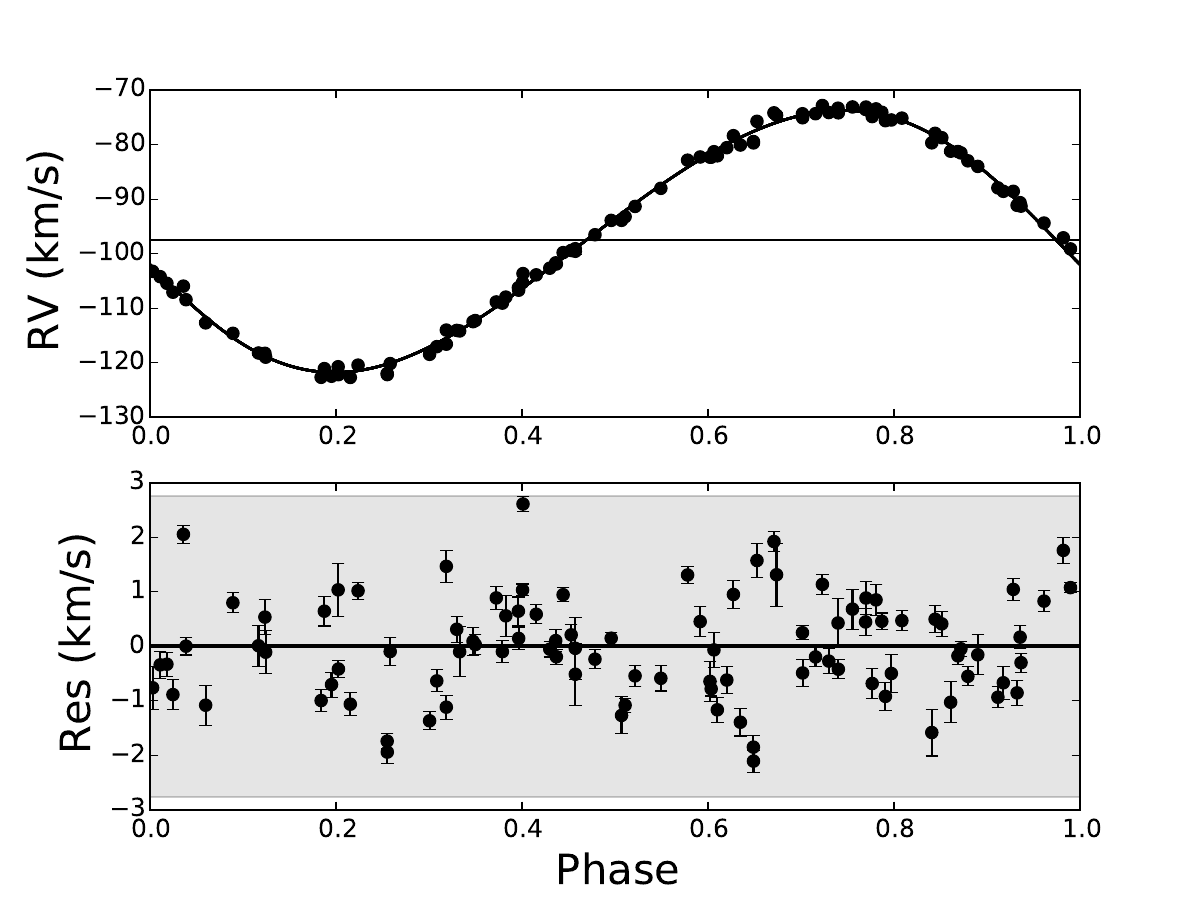}
\caption{Phase-folded radial velocities of the primary component of BD+46$\degr$442 (upper) with a period of 140.80 days and the residuals of the fit (lower) with the grey area as $\pm 3\sigma$ margin. The line curve in the upper panel denotes the RV curve for a Keplerian orbit with the best-fit orbital parameters given in Table~\ref{tab:bdorbpar}.}\label{fig:bdrvphase}
\end{figure}

Since we have a larger sample of 104 spectra available, we updated the radial velocities and orbital parameters by applying the same method.
The redetermined radial velocities are given in Table~\ref{tab:BDradvel} of the appendix, and they are plotted as a function of time and phase in Figs.~\ref{fig:rvtime} and \ref{fig:bdrvphase}, respectively. The orbital parameters obtained from the orbital fit solution are period, $P$, time of periastron passage, $T_0$, eccentricity, $e$, argument of periastron, $\omega$, systemic velocity, $\gamma$, semi-amplitude of the evolved component, $K_1$, semi-major axis of the evolved component, $a_1\sin i$, and the mass function $f(\mathrm{m})$. These orbital parameters are given in Table~\ref{tab:bdorbpar}.
 From here on, we consider the giant as the primary component and the companion as the secondary component.\\
\begin{table}
\begin{center}
\caption{Spectroscopic orbital solutions of the primary component of \bd.}
\label{tab:bdorbpar}
\begin{tabular}{l cc}\\
\hline
\hline 
Parameter & Value & $\sigma$ \\
\hline
$P$ (d) & 140.80& 0.03\\
$T_0$ (BJD) & 2455095.1682(6) & 1.7468(4)\\
$e$ & 0.076 & 0.006\\
$\omega$ ($^{\circ}$) & 100 & 4\\
$\gamma$ (km s$^{-1}$) & -97.44 & 0.10 \\
$K_1$ (km s$^{-1}$) & 24.02 & 0.15\\
$a_1\sin i$ $(\mathrm{AU})$ & 0.310 & 0.002\\
$f(\mathrm{m})$ $(M_\odot)$ & 0.200 & 0.004\\
\hline
\end{tabular}
\end{center}
\end{table}

\begin{figure*}[]
\centering
  \begin{tabular}{@{}c@{}c}
    \includegraphics[width=.4\linewidth]{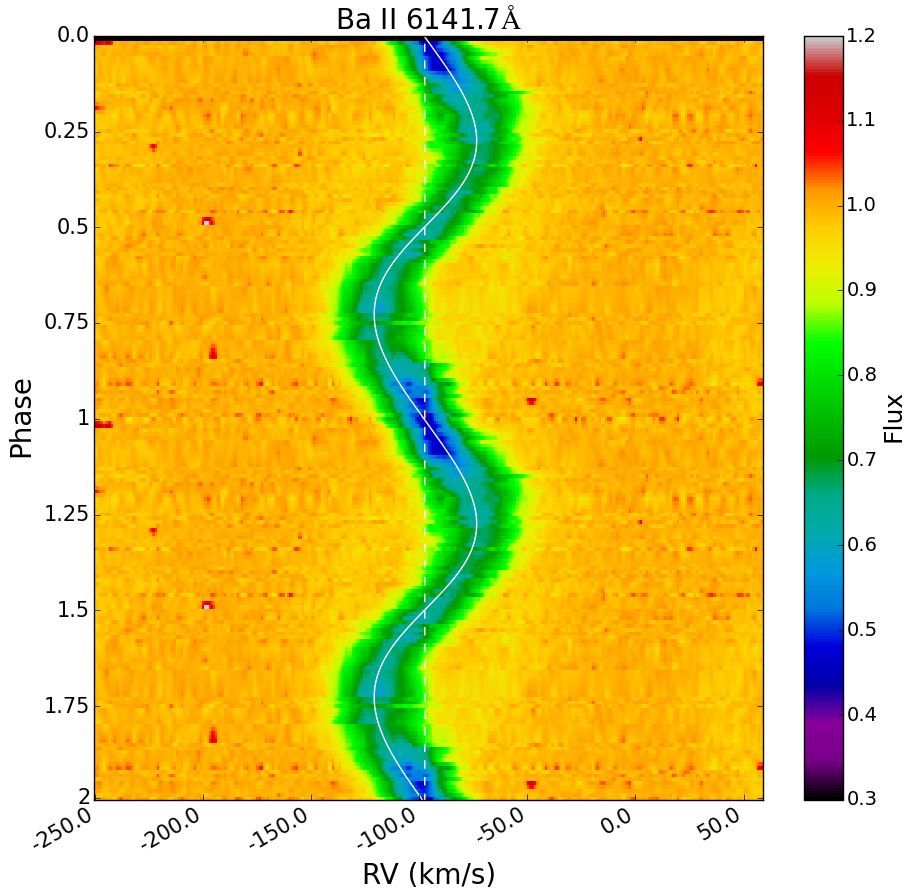} & 
    \includegraphics[width=.4\linewidth]{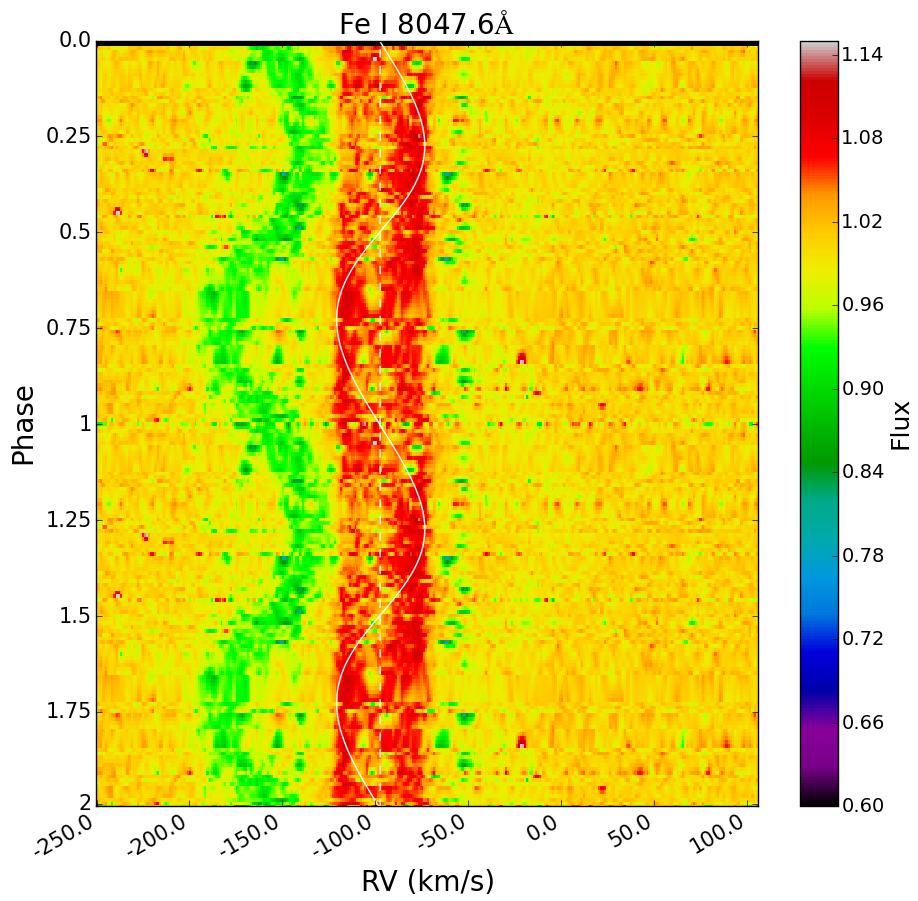} 
  \end{tabular}
  \begin{tabular}{@{}c@{} c}
    \includegraphics[width=.4\linewidth]{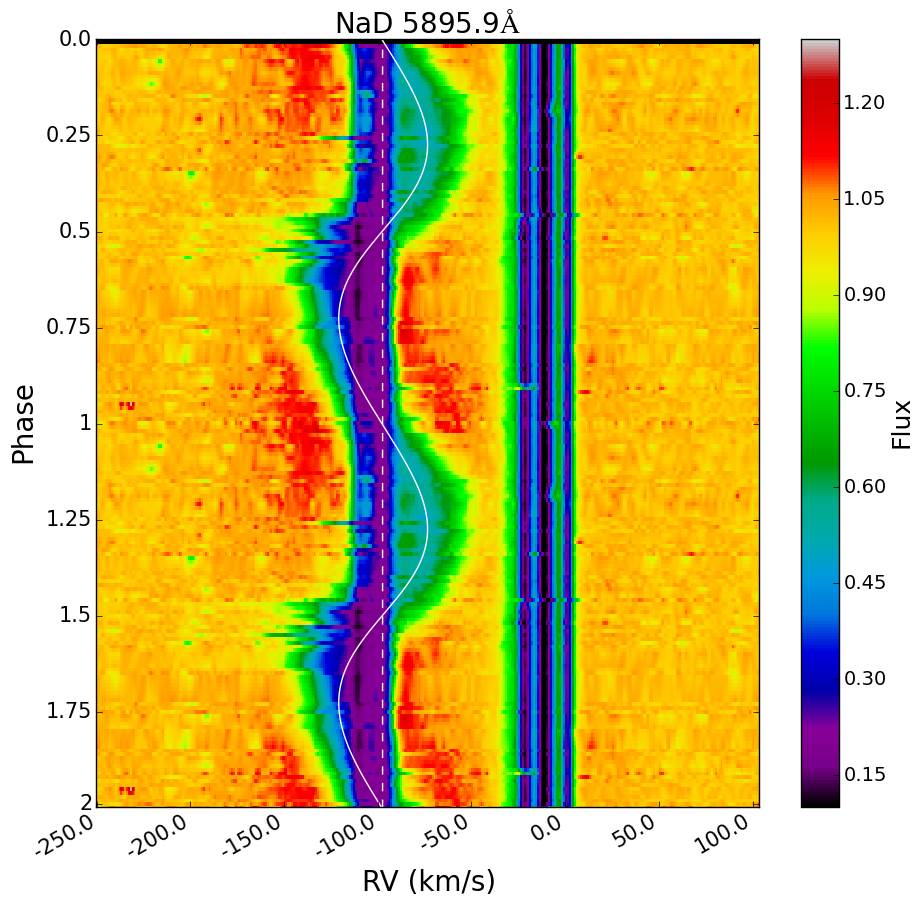} &
    \includegraphics[width=.4\linewidth]{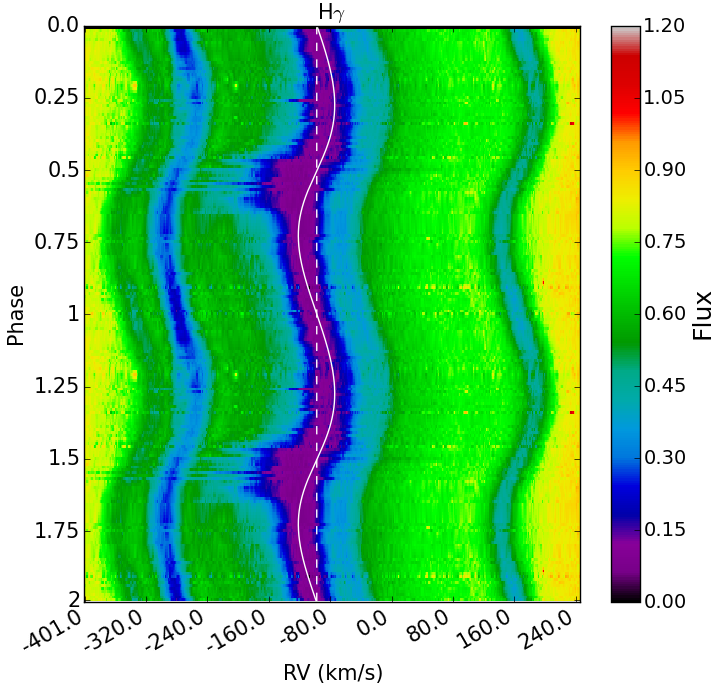} 
  \end{tabular}

  \begin{tabular}{@{}c@{} c}
    \includegraphics[width=.4\linewidth]{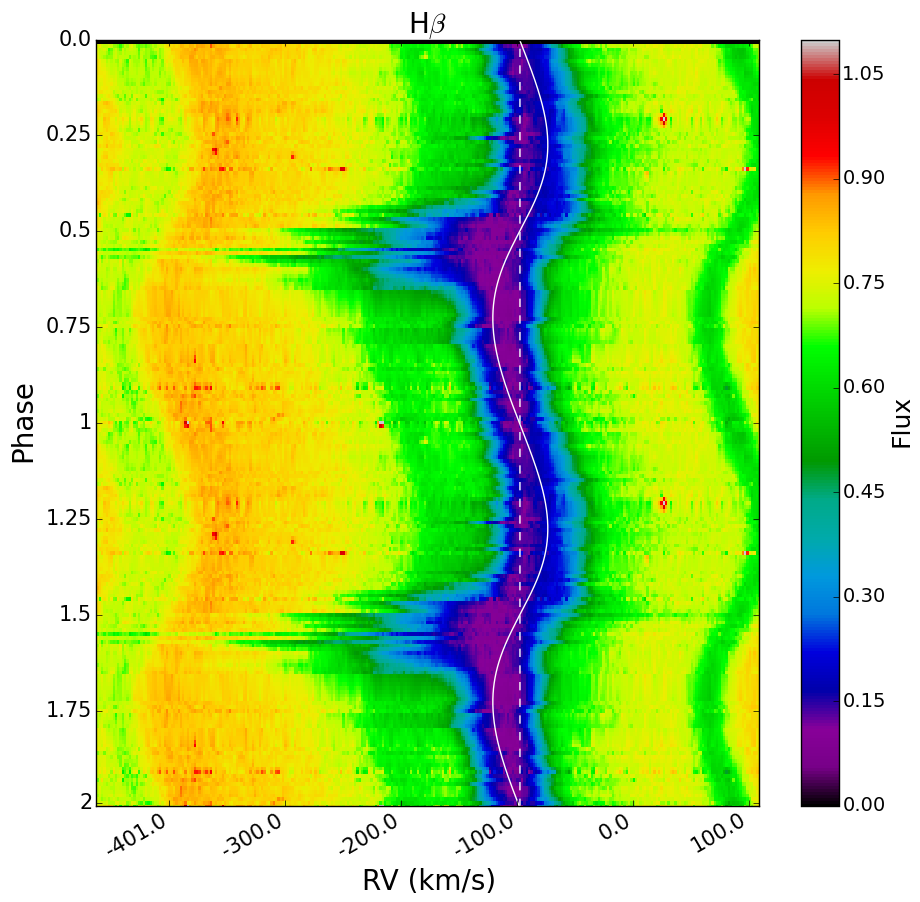} &
    \includegraphics[width=.4\linewidth]{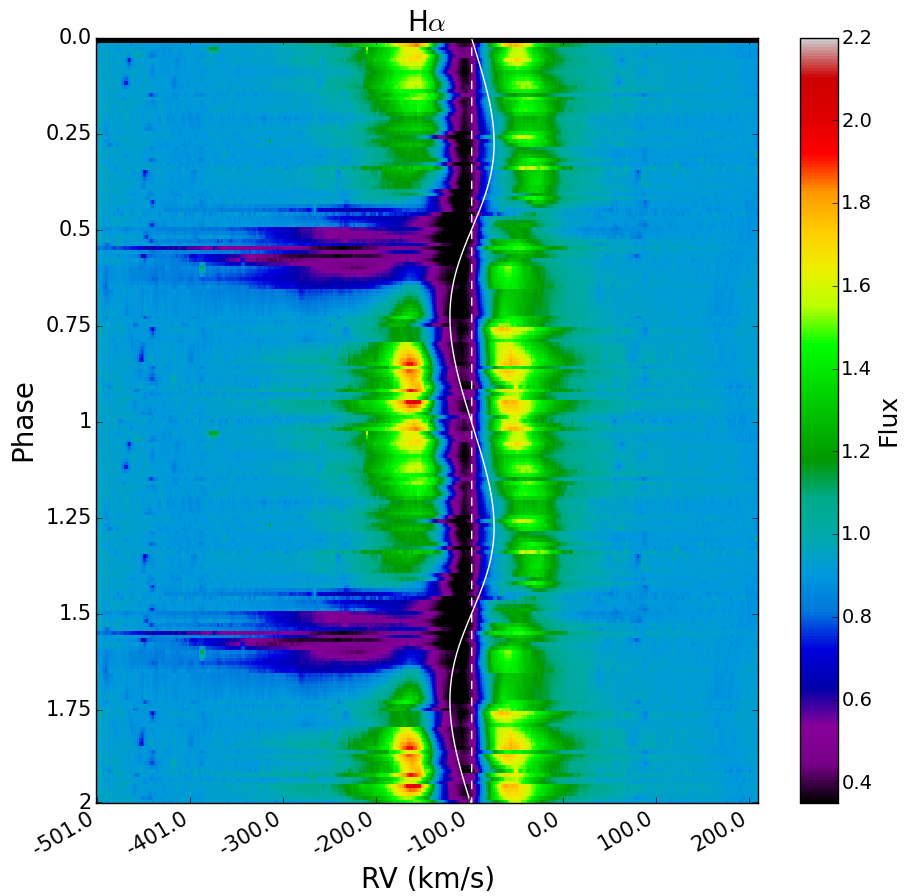} 
  \end{tabular}
  \caption{Dynamic spectra of Ba\,II, Fe\,I, NaD, H$\gamma$, H$\beta$, and H$\alpha$ of \bd\, as a function of orbital phase, showing different phase-dependent variations in the line profiles. The velocity on the x-axis denotes the RV shift from the laboratory wavelength. The white dashed line represents the systemic velocity, and the white full line represents the RV curve of the primary component. The colour indicates the continuum-normalised flux.}\label{fig:myfig}
\end{figure*}

We represent the phase-dependent variations of the spectral lines in the dynamic spectra of Fig.\ref{fig:myfig}.
%The phase-dependent variations in the spectral lines of \bd\, can be studied from the dynamic spectra shown in Fig. \ref{fig:myfig}. 
The Ba\,II 6141.7\,\angstrom\, line follows the RV curve of the primary component. Hence, this absorption line is a photospheric component of the primary (also present in other spectral lines not illustrated in this paper, e.g. Ca\,II, Fe\,II, NaD, Mg\,I, and K\,I).
\begin{figure*}[t!]
\centering
  \begin{tabular}{@{}c@{}c}
	\includegraphics[width=.5\textwidth]{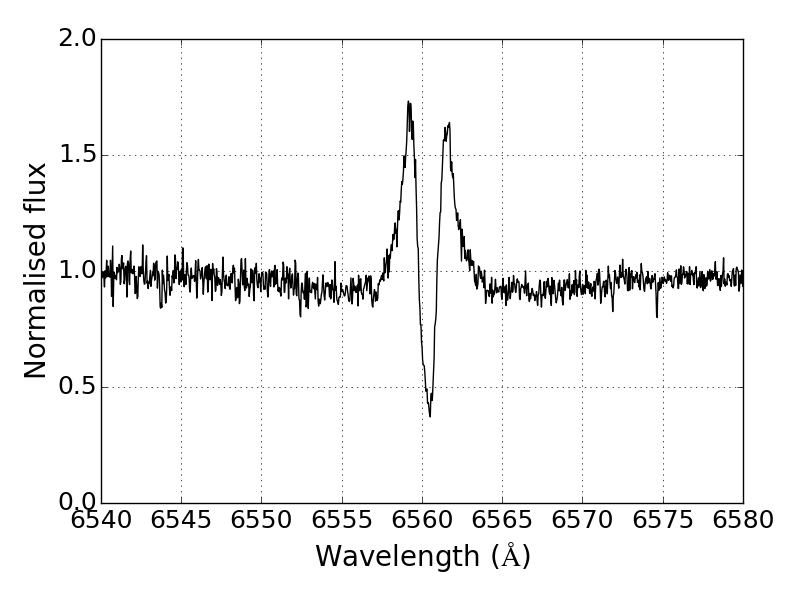}\label{fig:bdinfspec} &
	\includegraphics[width=.5\textwidth]{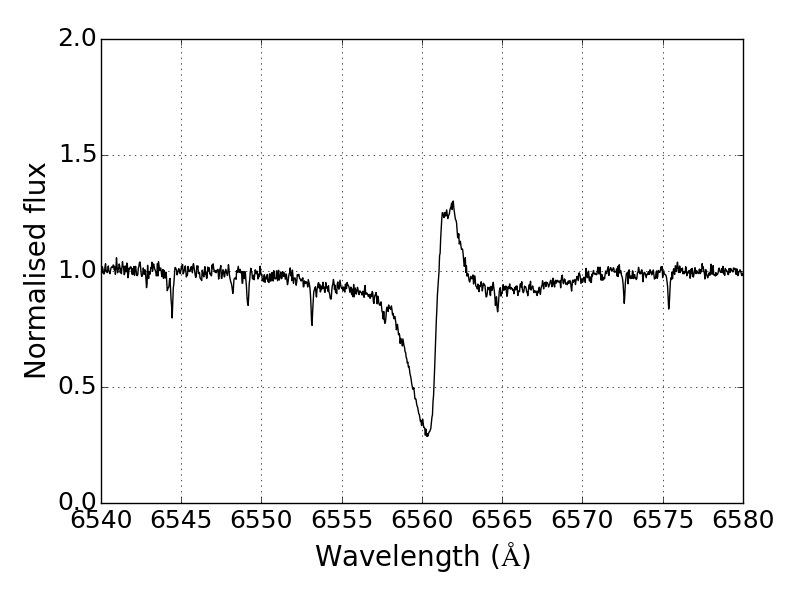}\label{fig:bdsupspec}
\end{tabular}
\caption{Continuum-normalised H$\alpha$ profile at inferior conjunction or at phase $\phi =0$ (left), and superior conjunction or at phase $\phi =0.5$ (right).}\label{fig:bdsupinfspec}
\end{figure*}
The Fe\,I 8047.6$\,\angstrom$ emission profile shows a constant RV throughout the orbital phase, with a FWHM $\approx 50\,$km s$^{-1}$. This emission most likely originates from a circumbinary gaseous disk that is located in between the binary and the dusty circumbinary disk \citep{gorlova12}. \citet{grundstrom07} proposed a similar gaseous disk to be created by a trailing gas stream in another system, namely RY Scuti. The gas stream flows through the L2 point and spirals around the system, from which a gaseous disk would eventually be established. Following this hypothesis, we assumed the gaseous disk around \bd\, to have an inner disk rim at a similar distance from the centre of mass of the binary system as the L2 point. For inclinations in the range of $40-90^\circ$, the L2 point will be located at $1.13-0.73\,$AU from the centre of mass. The HWHM ($\approx25\,$km s$^{-1}$) of the Fe\,I 8047.6$\,\angstrom$ is a good indication for the velocities of the matter in the inner regions of the disk. When we use the total mass of the binary system as the central mass, the radius corresponding to Keplerian velocities of $25\,\mathrm{km\,s}^{-1}$ ranges between $4.4-1.2\,$AU for inclinations of $40-90^\circ$. Hence, the radius of the inner disk region is indeed larger, but of the same order as the radius of the L2 point, as expected.

The most interesting variations appear in the dynamic spectra of H$\alpha$, H$\beta$, and H$\gamma$. These variations are due to a high-velocity outflow or jet that originates from an accretion disk around the companion, as discussed in Sect. \ref{sec:tsa}. \citet{gorlova12} performed no quantitative analysis. In this work, we provide a more thorough analysis on the origin and properties of the jet.

\section{Observations and data reduction}\label{sec:obs}
The system \bd\, is part of a large RV monitoring programme on evolved binaries, to study binary evolutionary channels \citep{vanwinckel10}. The observations were performed in order to achieve accurate RV estimates to derive the orbital parameters of the binary systems. One hundred and four spectra of \bd\, were obtained between July 2009 and January 2016. The high-resolution ($R\sim 85000$) optical spectra were obtained with the HERMES spectrograph \citep{raskin11}
 at the 1.2m Mercator telescope at the Roque de los Muchachos Observatory, La Palma, Spain. The HERMES spectrograph is located in a temperature-controlled room and is connected to the telescope by an optical fibre in order to optimize wavelength stability and efficiency. The observing dates and exposure times of the spectra are shown in Table \ref{tab:BDradvel} of the appendix. The exposure times of the spectra range between $500-1800\,$s, covering 55 orders and a wavelength range from 3770 to 9000$\,\angstrom$. A Th-Ar-Ne lamp was used as calibration source for the wavelength calibration. The Th-Ar-Ne exposures, biases, and flats were taken at the beginning and end of the night, with an additional wavelength calibration in the middle of the night. The raw spectra were reduced automatically at the end of each night using the specific Python-based HERMES pipeline. The reduction procedure of the pipeline also includes the correction for the barycentric motion. The RV stability is found to be as good as $60\,\mathrm{m\,s}^{-1}$ based on observations of more than 2000 RV standards over the whole time-line. The main contributor to the velocity zero-point shift is the pressure variation.

\section{Time-series analyses of the hydrogen Balmer lines}\label{sec:tsa}
The most pronounced variations in the dynamic spectra appear in the H$\alpha$ profile of \bd . This line profile varies between a double-peaked emission profile and a P Cygni-like profile (Fig. \ref{fig:bdsupinfspec}). A constant central absorption component is also present and blueshifted with $-13\,$km s$^{-1}$ from the systemic velocity. It is noteworthy that none of these features follows the RV curve of the primary, which indicates that they are unrelated to the photosphere of the primary. The double-peaked emission profile is most pronounced during inferior conjunction ($\phi =0$) and is centred on the central absorption component, from which the blueshifted and redshifted emission lines are shifted with $-60\,$km s$^{-1}$ and $60\,$km s$^{-1}$, respectively. Half an orbital phase later, at superior conjunction ($\phi = 0.5$), the companion is located between us and the primary, and the double-peaked emission profile has changed to a P Cygni-like profile with an extended blueshifted absorption wing. This wing reaches velocities of between 300 and $400\,$km s$^{-1}$ from the systemic velocity. A similar feature can be seen in the H$\beta$ and H$\gamma$ profile, and is slightly present in the NaD profile variations. It is interesting to note that the central absorption feature in the hydrogen lines decreases in strength from H$\alpha$ to H$\gamma$, unlike the photospheric component of the primary, which becomes more pronounced for higher Balmer lines.\\

\begin{figure*}[t]
\centering
\begin{tabular}{@{}c@{}c}
	\includegraphics[width=.5\textwidth]{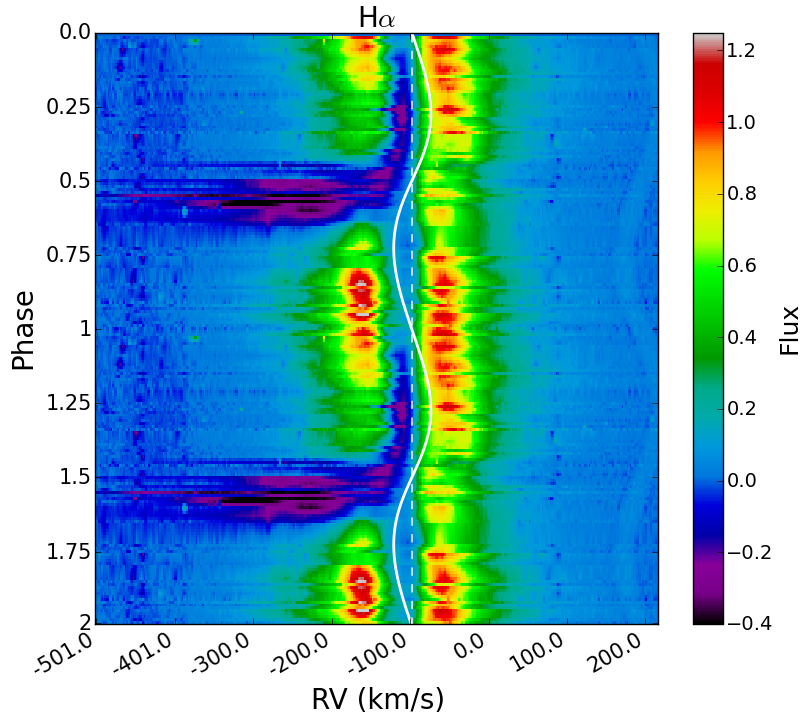} &
	\includegraphics[width=.5\linewidth]{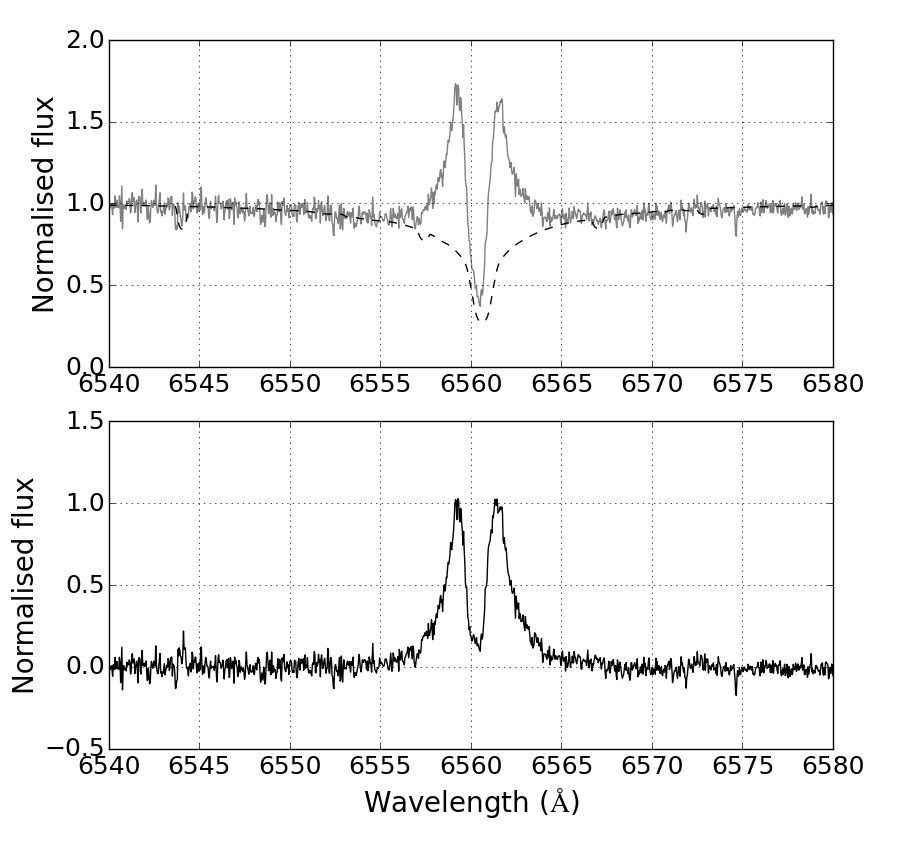}
\end{tabular}
\caption{\textit{Left}: Dynamic spectra of the photospheric-subtracted H$\alpha$ profile for \bd\, as a function of orbital phase. \textit{Right}: The original continuum-normalised H$\alpha$ profile at phase $\phi = 0.99$ (full grey line). The synthetic photospheric model (dashed black line) is overplotted with stellar parameters $T_{\mathrm{eff}} = 6250\mathrm{K}$, $\log g = 1.5$, and $[\mathrm{M/H}] = -0.5$. The lower right panel is the photospheric-subtracted profile.}  \label{fig:primrem}
\end{figure*}

Figure \ref{fig:primrem} shows the photospheric-subtracted dynamic spectra of H$\alpha$. A synthetic stellar spectrum of \citet{coelho05} \footnote{http://specmodels.iag.usp.br/fits\_search/?refer=s\_coelho05} was used for this substraction as the model spectrum of the primary, with stellar parameters $T_{\mathrm{eff}} = 6250\pm 250\mathrm{K}$, $\log g = 1.5\pm0.5$, and $[\mathrm{M/H}] = -0.7\pm0.2$.
Again, we can distinguish two main features in these dynamic spectra of H$\alpha$: (1) an extended blueshifted absorption wing, appearing between phase $\phi = 0.4$ and $\phi = 0.7$, (2) and a double-peaked emission feature with a peak-to-peak difference in RV of $100\,$km s$^{-1}$. \\
The variations in the dynamic spectra of the H$\alpha$ profile can be explained by the formation of a high-velocity outflow or jet that originates from an accretion disk around the companion \citep{gorlova12}. This circum-companion accretion disk is likely established by an active mass-transfer from the giant or from the gaseous disk. During superior conjunction ($\phi =0.5$), our line-of-sight towards the observer is obscured by the jet. Hence, the continuum photons originating from the primary are scattered out of the line-of-sight by the H-atoms in the jet, such that we see the H$_{\alpha}$ in absorption. The blueshifted absorption reaches radial velocities of $-400\,$km s$^{-1}$, and thus corresponds to the high velocities of the H-gas inside the jet.\\

As \citet{gorlova12} pointed out, the double-peaked emission feature in the H$\alpha$ profile is a good indication for a gas disk in the system. We showed the disk to be circumstellar instead of circumbinary by calculating the innermost radius of the disk. When we assume the disk to be circumbinary, the innermost radius cannot be less than the semi-major axis of the primary component. For a mass ratio of $q=0.6$ and projected semi-major axis of $a_1 \sin i = 0.31\,$AU, the projected rotational velocities above $60\,$km s$^{-1}$ would already be reached at radii smaller than the semi-major axis of the primary. Hence, we conclude that the disk cannot be circumbinary. It is more likely for the emission to originate from the accretion disk around the secondary component, in which case the accretion disk would be located well inside the Roche lobe of the secondary. Hence, the formation of the jet likely originates from this accretion disk around the secondary component.\\
The only inconsistency that arises is the absence of an RV shift in the double-peaked emission. We would expect the emission to follow the RV curve of the secondary with an amplitude of $\sim 14\,$km s$^{-1}$ (which would be the case for $q=0.6$). Instead, the emission has a constant RV and is centred on the central absorption feature, which has an RV shift of $-13\,$km s$^{-1}$ from the systemic velocity. Thus, either the mass ratio is even lower or the double peaked emission has its origin elsewhere. In the following sections, we discuss this more thoroughly.

\section{Jet geometry of \bd}\label{sec:jetgeom}
To deduce the geometry of the jet and estimate the inclination of the system, the jet angle, and the velocity profile of the jet, we used a similar approach as in \citet{thomas13}. They studied the geometry and velocity structure of the RR bipolar jet. The main difference between the RR and \bd\, is that the RR is observed almost edge-on, with an inclination of 85$\degr$, and is obscured by its circumbinary disk. Hence, in the RR, the central star is seen indirectly by scattering below and above the circumbinary disk. This is not the case for \bd, where the central star is seen directly.

\subsection{Geometrical model}\label{sec:geomod}
\begin{figure}[t!]
\centering
\includegraphics[width=.5\textwidth]{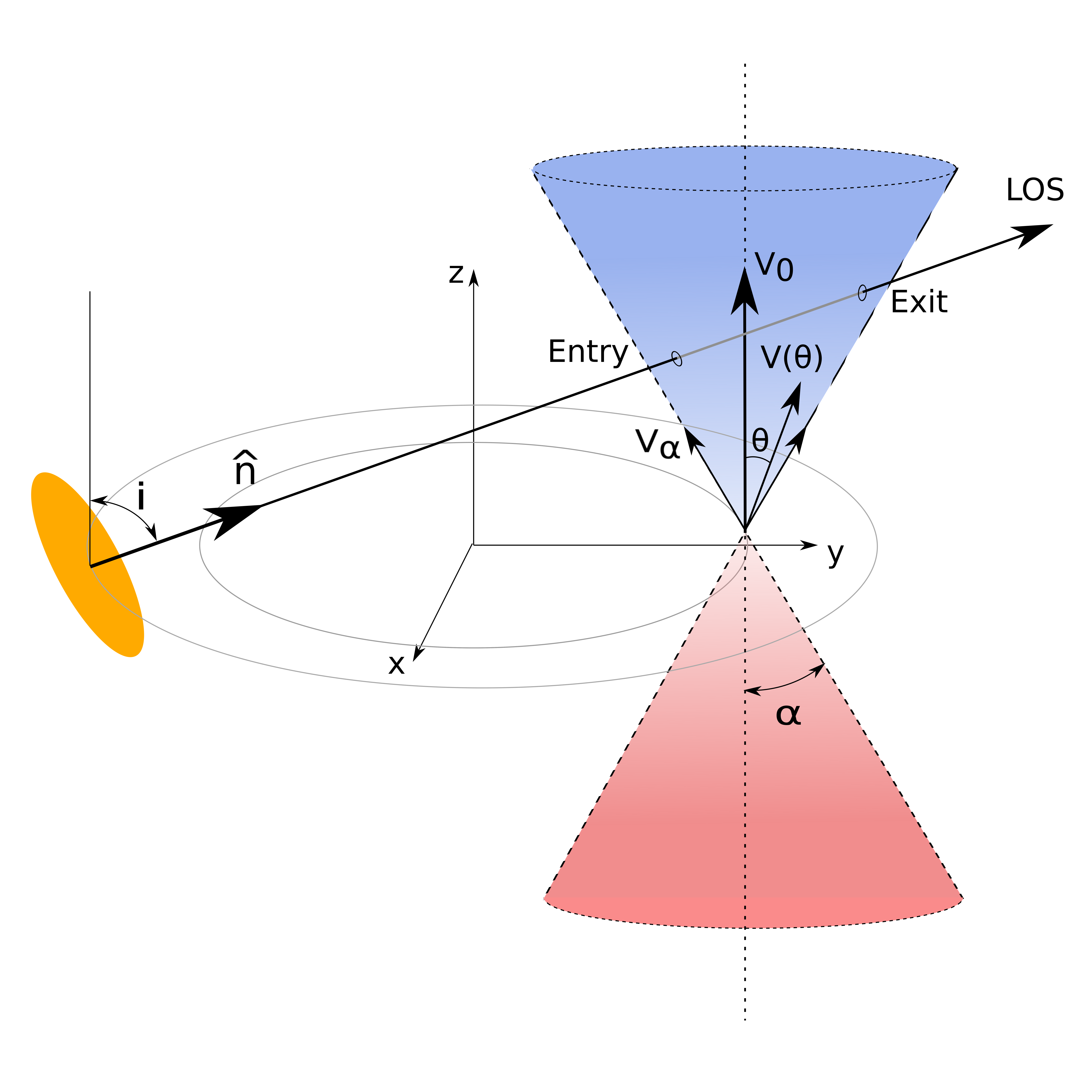}
\caption{Geometrical model for the calculations of jet angle and velocity profile. We assume the primary component to be a uniform disk, represented as the yellow ellipse in the illustration. This uniform disk is orientated normal to the line of sight and with its centre on the orbit of the primary. The bipolar outflow is considered as a double cone, which is centred at the orbital position of the secondary. The system depicted here is during superior conjunction or around phase $\phi = 0.5$.}\label{fig:geom}
\end{figure}

In order to determine the jet angle and the velocity profile of the jet, some assumptions were made for the geometrical model of the system. Figure \ref{fig:geom} illustrates this geometrical model adopted for the analytical calculations. A detailed description of these analytical calculations is presented in Appendix \ref{sec:A}. We show the primary component, which is represented as a uniform disk, and the bipolar outflow. This outflow originates from the secondary component and is considered as a double cone, with $\alpha$ being the half-opening angle of the cone. The axis of the cone is parallel with the z-axis. The disk and double cone are centred on the position of the primary (for the former) and the secondary (for the latter). The disk surface is normal to the unit vector $\hat{n}$, which represents the direction of the line of sight towards the observer. The line of sight always stays parallel to the yz-plane. The angle $i$ denotes the inclination of the system, and the angle $\theta$ is the latitudinal polar coordinate in the jet, which varies between $0$ and $\alpha$. The velocity at a certain position in the jet is given by $v(r,\theta)$, with $r$ the radial distance from the jet centre. Velocities at the jet axis are denoted by $v_0$, while those at the jet edges are given by $v_\alpha$.\\

To determine the orbit of the primary and secondary, we used the orbital parameters of Table~\ref{tab:bdorbpar}. The values for the semi-major axes of the orbits are determined by the applied inclination, which is a free parameter for the calculations. The radius of the primary (and thus also of the uniform disk in the model) is determined from the relation between luminosity, temperature, and stellar radius. We approximated the giant by a black body and applied the Stefan-Boltzmann law. Assuming a post-AGB luminosity of 2500$\,L_\odot$ and the effective temperature of $\teff = 6250\,$K, which was derived by \citet{gorlova12}, we obtain a radius for the primary of $R_p = 0.20\,$AU, which was applied in our model.

\subsection{Jet angle}\label{sec:jetangle}

The blueshifted absorption wing in Fig. \ref{fig:primrem} increases in strength from phase $\phi = 0.4$ to $\phi = 0.55$ and decreases again afterwards. This behaviour is correlated to the occultation of the primary by a high-velocity jet in our line of sight, and hence can be related to the amount of absorption by the jet. This absorption by the jet increases as the jet enters our line of sight, until the maximum coverage (during superior conjunction or $\phi =0.5$). Afterwards, this path length through the jet decreases again, and so does the amount of absorption by the jet.\\

In order to determine the inclination and jet angle, we related the amount of absorption in the H$\alpha$ profile with the amount of absorbing material in the jet. First, we determined the amount of absorption by the jet at a certain orbital phase by measuring the equivalent width $W_{\mathrm{H}\alpha}(\phi)$ of the H$\alpha$ profile. The amount of absorbing material can be represented by the column density of the jet $N_l$. Since we assumed the jet to be optically thin, the equivalent width is linearly proportional to the column density in the jet:
\begin{equation}
W = -N_l \int_{\nu_1}^{\nu_2}\sigma_\nu \dd \nu.
\end{equation}
Then, we compared the resulting equivalent width values of the observations with a model that calculates the orbital phase dependency of the normalised path length through the jet, for different values of inclination and jet angle.

\begin{figure}[t]
\begin{tabular}{@{}c}
	\includegraphics[width=.5\textwidth]{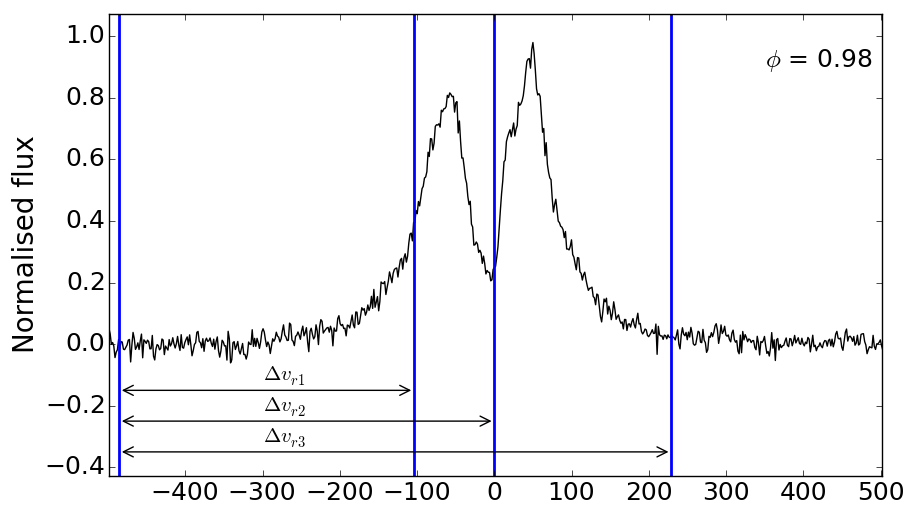} \\
	\includegraphics[width=1.\linewidth]{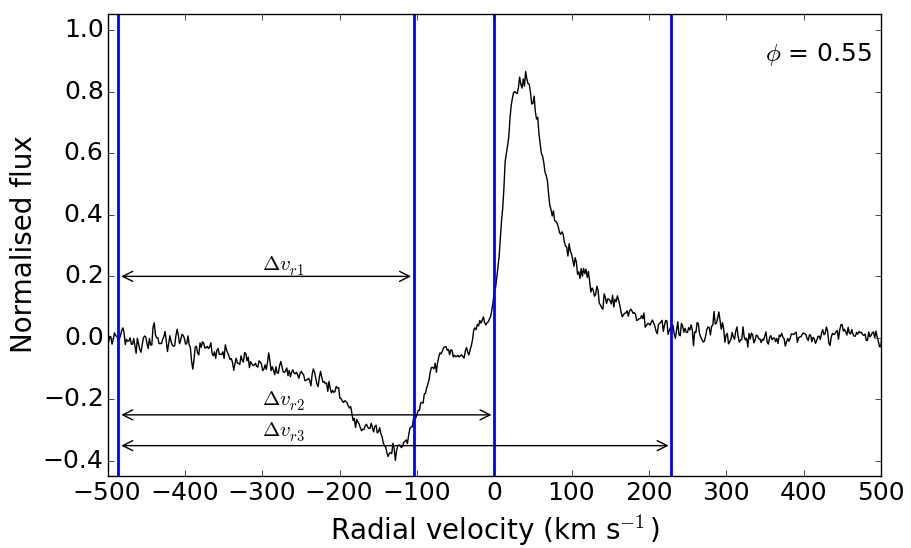}
\end{tabular}
\caption{Radial velocity ranges used for the equivalent width measurements of BD+46$\degr$442. The two spectra shown are at inferior conjunction or at phase $\phi =0$ (top), and superior conjunction or at phase $\phi =0.5$ (bottom). The three ranges $\Delta v_{r1}$, $\Delta v_{r2}$, and $\Delta v_{r3}$, are equivalent to the wavelength regions $\Delta \lambda_1$, $\Delta \lambda_2$, and $\Delta \lambda_3$.}  \label{fig:ewrangespec}
\end{figure}

The equivalent width $W_{\mathrm{H}\alpha}(\phi)$ of the H$\alpha$ profile was calculated for each photospheric-corrected spectrum in the following wavelength regions:
$\Delta \lambda_1 = \left[6550.00;6558.35\right] \angstrom$, $\Delta \lambda_2 = \left[6550.00;6560.63\right] \angstrom$, and $\Delta \lambda_3 = \left[6550.00;6565.63\right] \angstrom$,
which corresponds to a range in RV shift from the systemic velocity of
$\Delta v_{r1} =   \left[-486;-104\right]\,$km s$^{-1}$, $\Delta v_{r2} =   \left[-486;0\right]\,$km s$^{-1}$, and $
\Delta v_{r3} =   \left[-486;229\right]\,$km s$^{-1}$. These three ranges are also shown in Fig. \ref{fig:ewrangespec}.
The equivalent width was calculated for three different regions instead of just one, since each region includes or excludes different features of the H$\alpha$ profile.

\begin{figure}[t]
\centering
\includegraphics[width=.5\textwidth]{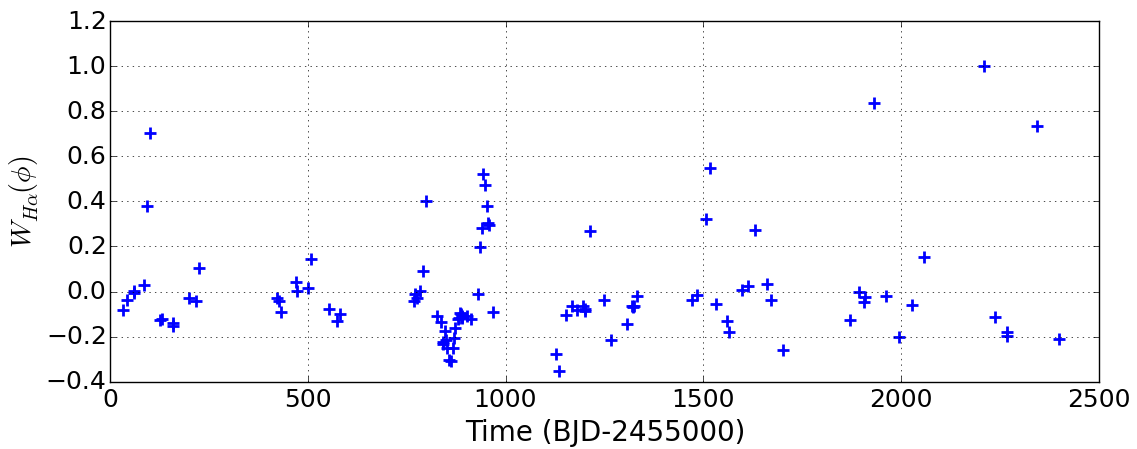}
\caption{Equivalent width measurements (in units of $\angstrom$) for BD+46$\degr$442 in the wavelength region $\Delta \lambda_1$ for each observation.}  \label{fig:ew1time}
\end{figure}

The variations in equivalent width are phase-dependent and reach a maximum value during the occultation by the jet. However, the maximum value of the equivalent width is different for each cycle of 140.8 days, as shown in Fig. \ref{fig:ew1time}. 
This is an indication that the outflow itself is variable throughout time and suggests that the mass-transfer rate is variable as well. A similar variable mass-transfer rate has been suggested for cataclysmic variables \citep{knigge11}, where the variable mass transfer can be caused by the expansion and contraction of the donor star, which can affect the evolution of accreting white dwarf systems \citep{toonen14a}.
\begin{figure}[t]
\centering
\includegraphics[width=.5\textwidth]{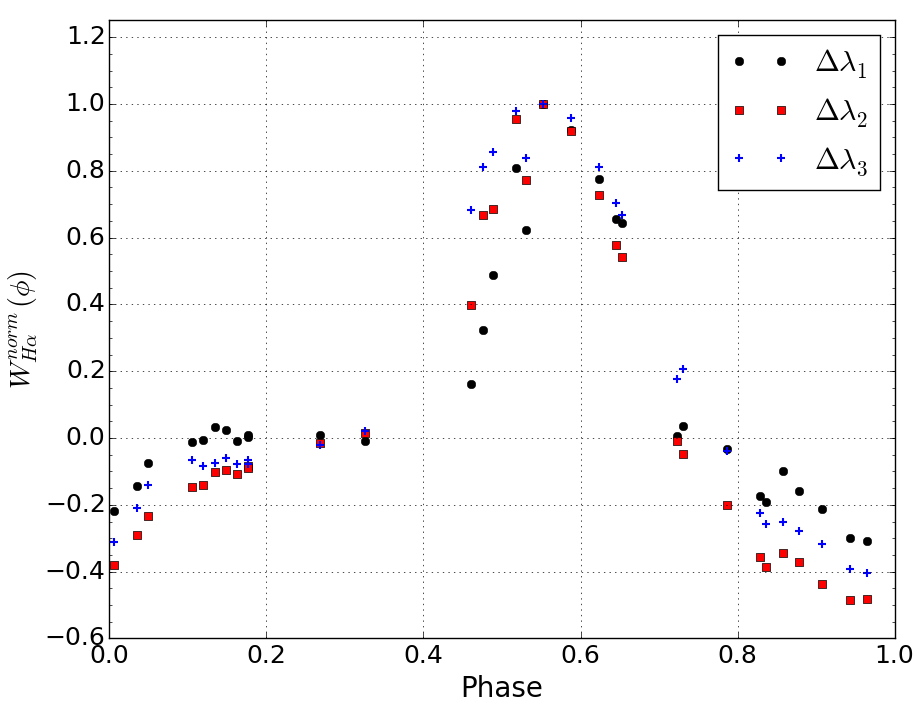}
\caption{Normalised equivalent width measurements for BD+46$\degr$442 of the observations taken between BJD = 2455790$\,$d and BJD = 2456000$\,$d. The circles, squares, and crosses represent the measurements of the equivalent width for the wavelength ranges $\Delta\lambda_1$, $\Delta\lambda_2$, and $\Delta\lambda_3$ respectively.}  \label{fig:ewnormall}
\end{figure} \\

\begin{figure}[t]
\centering
\includegraphics[width=.5\textwidth]{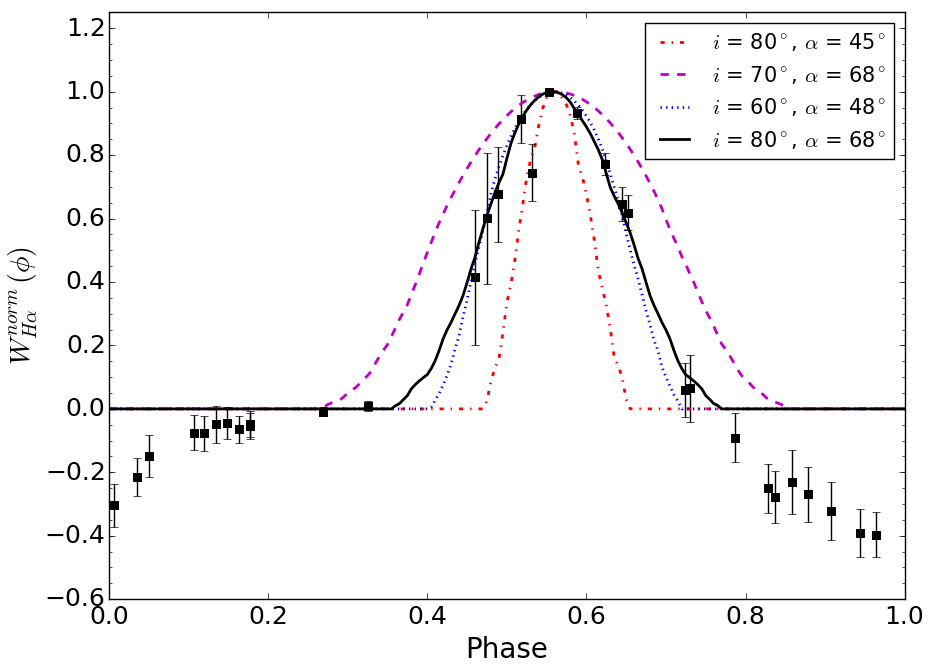}
\caption{Normalised equivalent width measurements for BD+46$\degr$442 of the observations taken between BJD = 2455790$\,$d and BJD = 2456000$\,$d. Each data point is the mean of the measurements for the three wavelength ranges, and the error bars represent one standard deviation. Model calculations (dash-dot, dotted and dashed curves) and the best-fit model (solid black curve) are overplotted. }  \label{fig:ewfit}
\end{figure}

Since the equivalent width measurements at different orbital cycles are variable, we only considered the observations taken between the time interval BJD = $2455790-2456000\,$d, which is well sampled. The normalised equivalent width measurements for these observations were compared with a model that determines the normalised path length $s(\phi)$ through the jet at each orbital phase, shown in Fig. \ref{fig:ewnormall}. 
All data points are scaled according to the minimum and maximum values:
\begin{equation}
W_{\mathrm{H}\alpha}^\mathrm{norm} (\phi)= \frac{W_{\mathrm{H}\alpha}(\phi)-W_\mathrm{max}}{W_{\mathrm{max}}-W_\mathrm{min}}. \label{eq:ewscaling}
\end{equation} 
As we can see in Fig. \ref{fig:ewfit}, the equivalent width only stays constant between phases $\phi = 0.1$ and $\phi = 0.4$, but drops to its lowest value around phase $\phi \approx 0$. If the change in equivalent width would only depend on the absorption by the jet, however, we would expect it to stay constant at a minimum value, when the primary is not obscured by the jet. Since multiple features in the system can affect the equivalent width measurements, this decrease is most likely caused by an additional source of emission, such as a hot spot in the accretion disk. Hence, the mean of the measurements at phases $\phi = 0.27$ and $\phi = 0.33$ was taken as minimum value. The normalised path length was fitted with the normalised equivalent width measurements using a $\chi^2$-test. The equivalent width measurements between orbital phases $\phi = 0-0.2$ and $\phi = 0.75-1$ were not used for the fitting procedure, since they include effects from other features in the system. The phase of maximum absorption of the model was allowed to be variable.

\begin{figure}[t!]
\centering
\includegraphics[width=.475\textwidth]{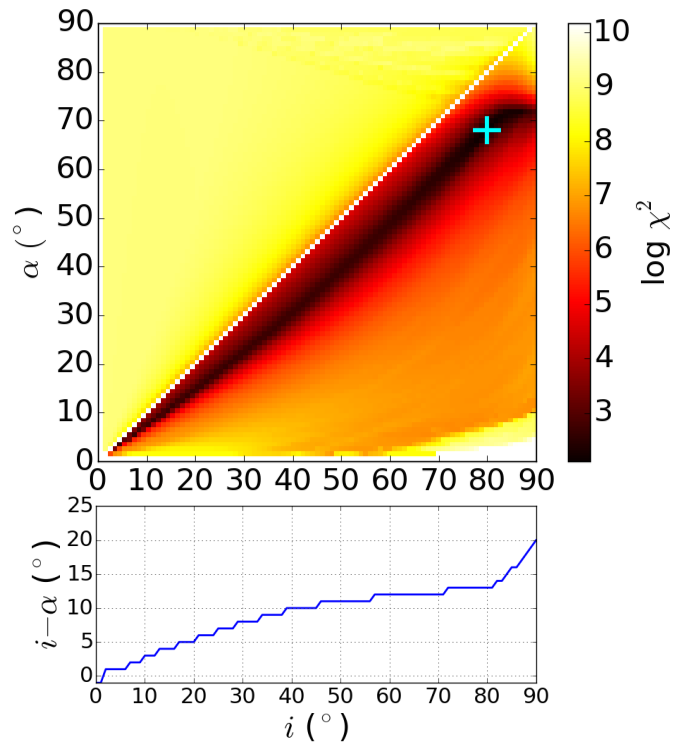}
\caption{$\log\chi^2$ results for inclination and jet angle. The cross indicates the best-fit model with parameters $i=80\degr$ and $\alpha = 68\degr$. The lower panel shows the difference between inclination and jet angle of the best-fit model for each value of inclination.}  \label{fig:chi2}
\end{figure}

The curve with parameters $i = 80\degr$ and $\alpha = 68 \degr$ shows the best-fit model, which matches the data fairly well. The model calculations also show the effect of a change in inclination $i$ or jet angle $\alpha$ on the curve. We note that by changing both $i$ and $\alpha$, but keeping the difference between the two angles constant (as is the case for the solid black curve and the dotted blue curve), the shape of the curve varies little and still fits the data points well. This can also be seen in the two-dimensional $\log\chi^2$-plot of the fitting (Fig. \ref{fig:chi2}). Hence, we do not assume the best-fit model as the exact solution for this system, but show that there is a clear correlation between the two angles.
The $\log\chi^2$-valley shows a linear relation between the inclination and jet angle. For inclinations between $i = 40\degr$ and $i = 80\degr$, the difference between inclination and jet angle is about $10-13\degr$.\\

\subsection{Velocity profile of the jet}\label{sec:jetvel}
The shape of the absorption wing in the H$\alpha$-profile does not only depend on the inclination and jet angle, but also on the velocity structure in the jet. Two possible velocity profiles for the jet are compared with the observed H$\alpha$ absorption feature. The first considered model has a constant velocity $v_0$ throughout the jet, with the velocity vector always pointing radially away from  the secondary component. The second model has a latitudinally dependent velocity structure, which is also independent of the radial component:
\begin{equation}
v(\theta) \hat{r} = \left[v_0+(v_\alpha-v_0)\left(\frac{\theta}{\alpha}\right)^p\right]\hat{r},\label{eq:vtheta}
\end{equation}
with $\theta$ the polar latitudinal coordinate and $v_\alpha$ the velocity at the jet edges. For the constant velocity model we have $v_{\alpha} = v_0$.

The latitudinally dependent velocity structure was suggested by \citet{thomas13} for the velocity structure of the outflow in the RR and considers the material in the jet with the highest velocities to be along the jet axis, while at the edges of the jet, the outflowing material has the lowest velocities. This profile, with an inner high-velocity component and an outer low-velocity component in the jet, is also observed in several accreting T Tauri stars \citep{coffey04,bacciotti00} and is similar to the two-component outflow described by \citet{romanova09}.
\begin{figure}[t]
\centering
\includegraphics[width =.5\textwidth]{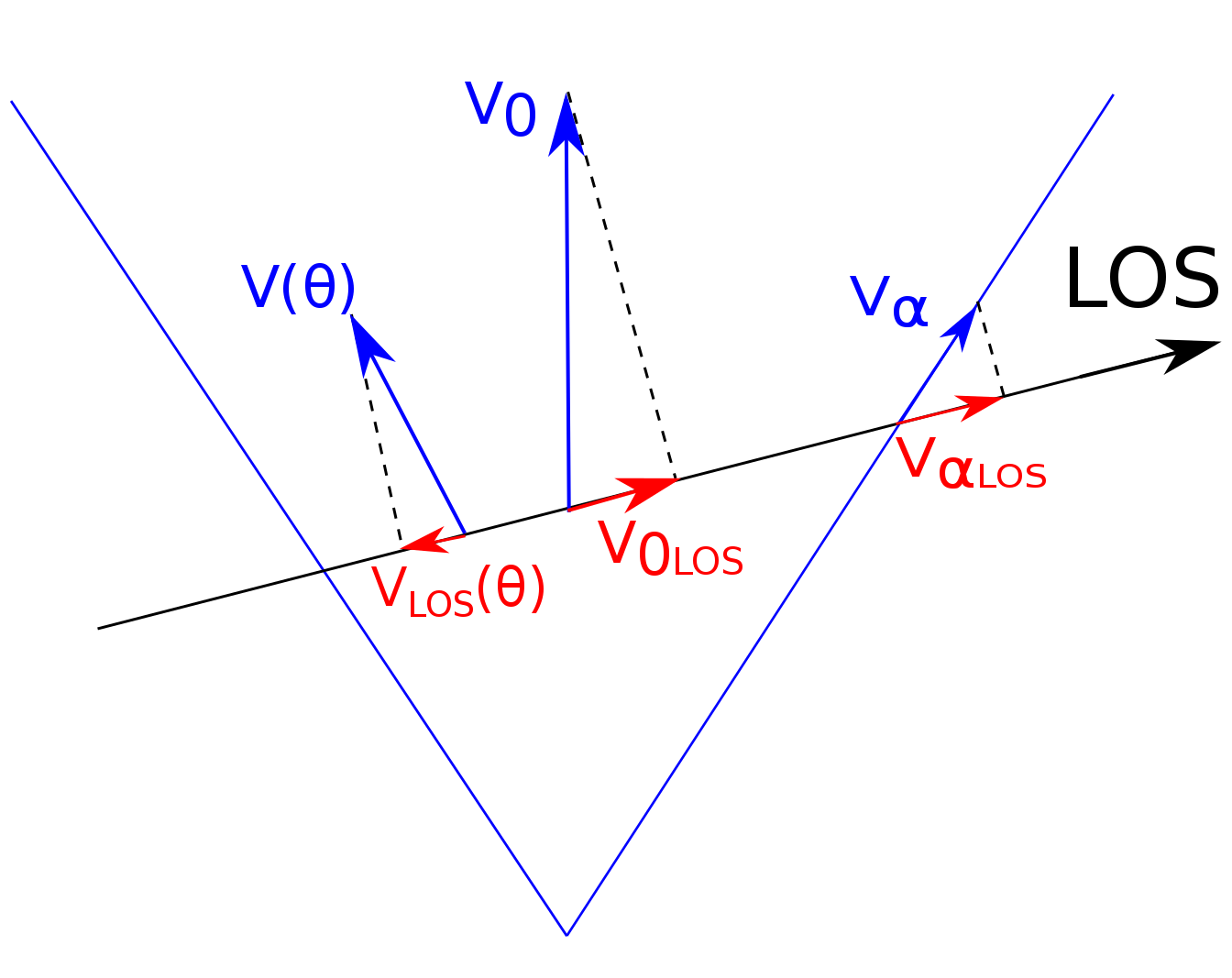}
\caption{Projection of the velocity in the jet onto the line-of-sight.}\label{fig:los}
\end{figure}

In order to determine theRV of the material in the jet towards the observer, we must project its velocity onto our line of sight (Fig. \ref{fig:los}):
\begin{equation}
v_\mathrm{LOS}(\theta) = v(\theta) \hat{r}\cdot \hat{n}-v_s(\theta),
\end{equation}
with $v_s(\theta)$ the RV component of the secondary.
These projected line-of-sight velocities were calculated as a function of orbital phase at three positions in the jet: the projected velocity at the jet entrance $v_{\alpha_1}^\mathrm{LOS} = v_{\alpha_1}\hat{r}_{\alpha_1}\cdot \hat{n}-v_s$, the projected velocity at the jet exit $v_{\alpha_2}^\mathrm{LOS} = v_{\alpha_2}\hat{r}_{\alpha_2}\cdot \hat{n}-v_s$, and the projected velocity of the intersection point of the line of sight with the plane parallel to the xz-plane, which contains the jet axis  $v_{0}^\mathrm{LOS} = v_{0}\hat{r}_{0}\cdot \hat{n}-v_s$.

The three projected velocities were calculated for both velocity structures. The geometrical model used for the calculations has the orbital parameters of Table \ref{tab:bdorbpar}. In Section \ref{sec:jetangle} we showed the relation between the inclination of the system and jet angle ($i-\alpha \approx 12\degr$), although neither angle could be determined precisely. For this reason, we used an example case for the geometrical model, which holds to the relation $i= 52\degr$ and $\alpha = 40\degr$.

\begin{figure*}[]
\captionsetup{width=1.0\textwidth}
\centering
  \begin{tabular}{@{}c}
\includegraphics[width=0.88\textwidth]{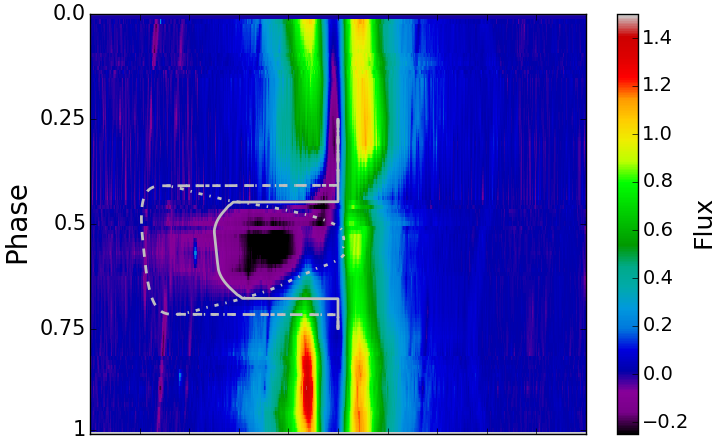}\label{fig:v0prim}\\
\includegraphics[width = 0.90\textwidth]{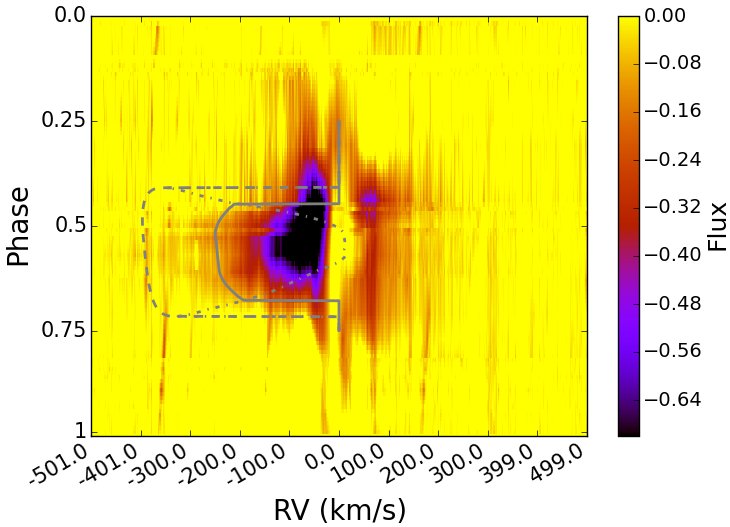}\label{fig:v0abs}
  \end{tabular}
\caption{Projected line-of-sight velocity for the constant velocity profile. These are plotted over the normalised dynamic spectra of the H$\alpha$ profile of BD+46$\degr442$ observed between BJD = 2455790$\,$d and BJD = 2456000$\,$d. The upper panel shows the observed spectra corrected for the photospheric contribution of the primary. The lower panel is the difference between the photospheric corrected spectra and a template spectrum taken at phase $\phi = 0.09$, representing the variations in the strength of the absorption feature as a function of orbital phase. The continuum corresponds to a flux value of zero. Hence, absorption is represented by negative flux values and emission by positive values. The grey curves correspond to $v_{0}^\mathrm{LOS}$ (full line), $v_{\alpha_1}^\mathrm{LOS}$ (dash-dotted line), and $v_{\alpha_2}^\mathrm{LOS}$ (dashed line).}  \label{fig:v0result}
\end{figure*}

\begin{itemize}
\item[\textbullet] \textbf{Constant velocity profile}

%$\mathrm{ }$\\
The constant velocity structure has a jet velocity of $v_0=400\,$km s$^{-1}$. This velocity $v_0$ was chosen such that the highest projected velocities match the highest velocity of the blue edge of the absorption feature in the H$\alpha$ profile at $-400\,$km s$^{-1}$. Figure \ref{fig:v0result} shows the projected line-of-sight velocities of the constant velocity profile as a function of orbital phase. These are plotted over the dynamic spectra of the H$\alpha$ profile. As in the previous section, we only considered the observation between the time interval BJD = $2455790-2456000\,$d, which is well sampled. Again, the reason not to include all observations is the variability of the jet outflow throughout time. \\%This causes the velocities in the jet to be variable as well. \\

The constant velocity model predicts the highest projected velocities to be reached close to the jet exit, $v_{\alpha_2}^\mathrm{LOS}$. According to the model, $v_{\alpha_2}^\mathrm{LOS}$ stays nearly constant during the obscuration of the jet. This disagrees with the RV of the observed blue edge of the absorption wing. From phase $\phi \approx 0.45$ to $\phi \approx 0.55$, this maximum blueshifted velocity of the absorption rises steeply to a maximum. From $\phi \approx 0.55$ to $\phi \approx 0.65$, the maximum blueshifted velocity decreases again until the absorption wing disappears.

Secondly, the strong absorption at $\sim -50\,$km s$^{-1}$ in the lower panel of Fig. \ref{fig:v0result} cannot be explained by this model. Between phases $\phi \approx 0.5$ and $\phi \approx 0.65$, this strong absorption region is located in between projected velocities $v_{\alpha_1}^\mathrm{LOS}$ and $v_{0}^\mathrm{LOS}$, where $v_{0}^\mathrm{LOS}$ appears to follow the edge of this absorption region. This would suggest that most of the absorption takes place close to the jet entrance, which is not the case, since the strongest absorption region does not follow the $v_{\alpha_1}^\mathrm{LOS}$-curve, but instead has a constant RV shift of $-50\,$km s$^{-1}$. These inconsistencies show that a constant velocity structure is unable to predict the observed absorption features.
\end{itemize}
\begin{figure*}[t!]
\captionsetup{width=1.\textwidth}
\centering
  \begin{tabular}{@{}c}
\includegraphics[width=0.90\textwidth]{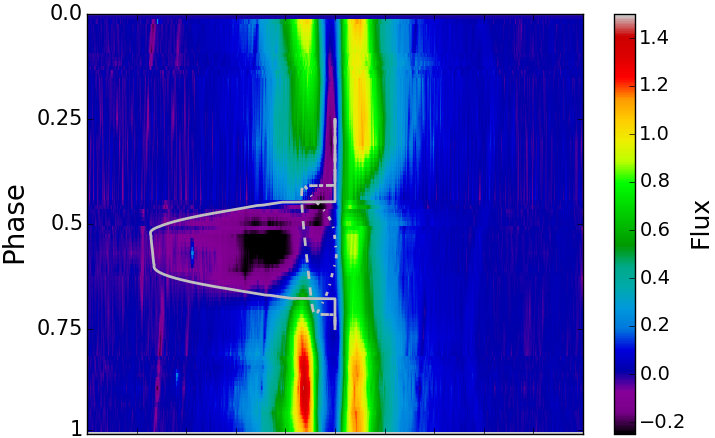}\label{fig:v2prim}\\
\includegraphics[width = 0.92\textwidth]{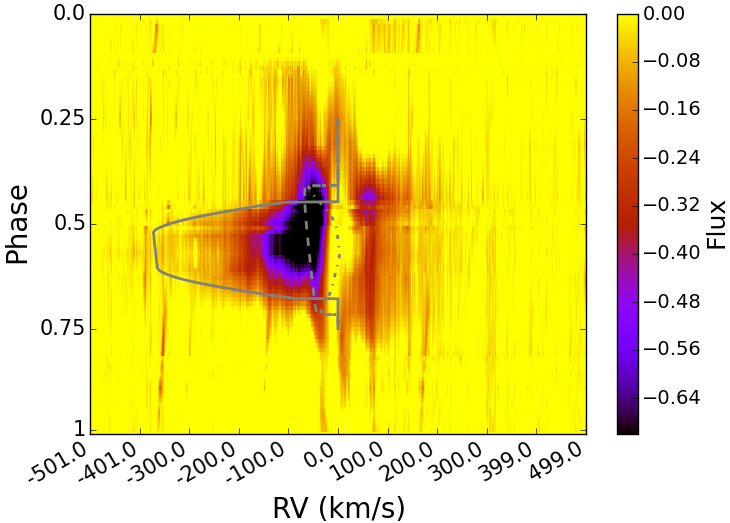}\label{fig:v2abs}
  \end{tabular}
\caption{Similar to Fig. \ref{fig:v0result}, but for the latitudinally dependent velocity profile. The grey curves correspond to $v_{0}^\mathrm{LOS}$ (full line), $v_{\alpha_1}^\mathrm{LOS}$ (dash-dot line), and $v_{\alpha_2}^\mathrm{LOS}$ (dashed line).}  \label{fig:v2result}
\end{figure*}
\begin{figure*}[t]
\captionsetup{width=1.\textwidth}
\centering
  \begin{tabular}{@{}c@{}c}
  \includegraphics[width =.5\textwidth]{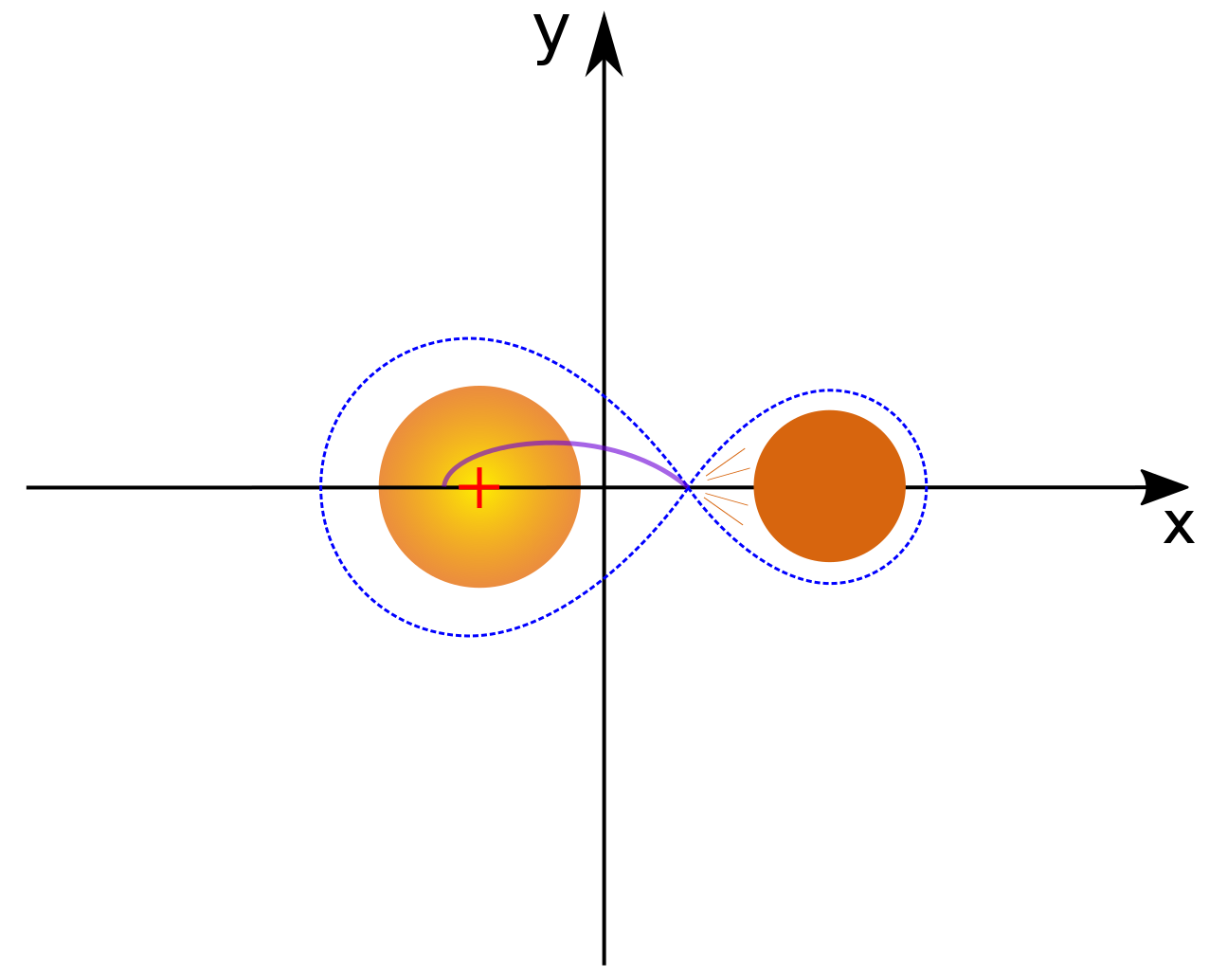}
%\caption{Cartesian coordinates}\label{fig:geomcart} 
&
	\includegraphics[width = .5\textwidth]{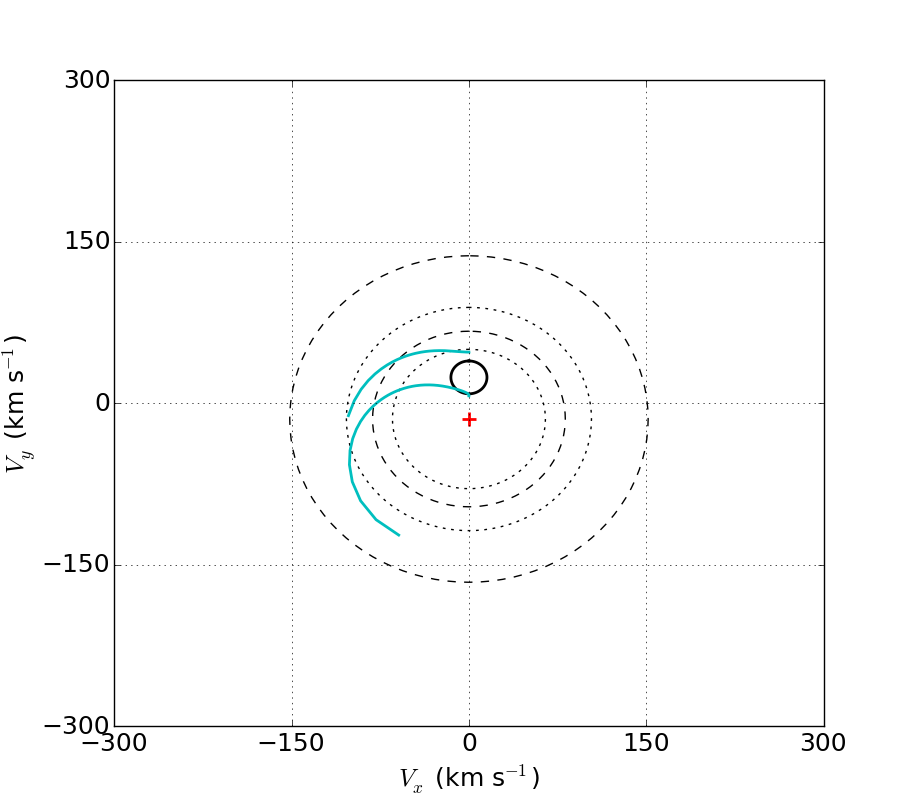}
\end{tabular}
\caption{\textit{Left}: Illustration of the assumed geometry of \bd\, for the Doppler tomography in the co-rotating frame of the binary. The mass centre of the binary coincides with the origin, and the blue dashed curve are the Roche lobes of the binary. The primary is located on the positive x-axis, and the position of the secondary is indicated with the cross on the negative x-axis. The purple curve represents the gas stream from the L1-point towards the secondary. The accretion disk is centred on the secondary. \textit{Right}: The system represented in velocity coordinates. The primary is represented as the full black circle, located on the positive $V_y$-axis. The location of the secondary is indicated by the cross on the negative $V_y$-axis. The curve originating from the primary indicates the actual velocity of the mass stream. The curve that originates at a higher $V_y$-velocity indicates the Keplerian velocities at the path of the mass stream. The velocities of the accretion disk are represented by the dashed and dotted circles, centred on the secondary. The four circles represent, from outside to inside, a projected Keplerian velocity of 150, 100, 75, and $50\,$km s$^{-1}$, respectively.}\label{fig:geomdopmap}
\end{figure*}

\begin{itemize}
\item[\textbullet] \textbf{Latitudinally dependent velocity profile}
%$\mathrm{ }$\\

The second profile is the latitudinally dependent velocity profile, where the velocities in the jet are given by Eq. \ref{eq:vtheta}, with $v_\alpha = 60\,$km s$^{-1}$ and $v_0 = 600\,$km s$^{-1}$. The projected line-of-sight velocities for the latitudinally dependent velocity structure,  $v_{\alpha_1}^{\mathrm{LOS}}$ (dash-dotted line), $v_{\alpha_2}^\mathrm{LOS}$ (dashed line), and $v_0^{\mathrm{LOS}}$ (full line) are plotted over the same dynamic spectra in Fig. \ref{fig:v2result}. We immediately notice the difference between the RV curves of both profiles. For the constant velocity profile, the highest projected velocities were reached at the jet axis, while the lowest radial velocities are reached at the jet entrance. This is different for the latitudinally dependent profile, where the highest projected velocities are reached along the plane that contains the jet axis and which is parallel to the xz-plane.\\

We note for this profile that the projected velocity curve of $v_{0}^\mathrm{LOS}$ matches the blue edge of the absorption wing fairly well. This suggests that the absorption, which is observed at the highest blueshifted velocities, occurs in the plane in the jet that contains the jet axis and which is parallel to the xz-plane. \\
The projected velocities at jet entrance $v_{\alpha_1}^\mathrm{LOS}$ and jet exit $v_{\alpha_2}^\mathrm{LOS}$ stay more constant throughout the orbital phase. $v_{\alpha_2}^\mathrm{LOS}$ only shows a small shift in velocity that is due to the orbital motion of the secondary. Moreover, $v_{\alpha_2}^\mathrm{LOS}$ matches the absorption maximum of the dynamic spectra in the lower panel of Fig. \ref{fig:v2result} very well. The absorption maximum appears around phase $\phi \approx 0.4$ at an RV shift of $\sim -50\,$km s$^{-1}$. At this phase, the line of sight enters the jet, and thus the absorption is caused by the material at the edge of the jet. This is in agreement with $v_{\alpha_1}^\mathrm{LOS}$ and $v_{\alpha_2}^\mathrm{LOS}$ of the model, which represent the velocities at the edges and are also first seen at phase $\phi = 0.4$. $v_{\alpha_2}^\mathrm{LOS}$ seems to follow the absorption maximum of the H$\alpha$ profile reasonably well. This is not the case for $v_{\alpha_1}^\mathrm{LOS}$, however, since it only matches the absorption maximum around phases $\phi = 0.4$ and $\phi = 0.7$.
\end{itemize}
\subsection{Discussion and conclusions on the geometry of \bd}\label{sec:concgeo}
%***Bij sectie \textit{Conclusions}?***\\
We conclude that the inclination or jet angle cannot be determined without constraints on the inclination, as there is a clear correlation between the two parameters. Thus, if the inclination is known, the jet angle can be determined as well. By comparing the latitudinally dependent velocity structure with the H$\alpha$ absorption feature, we conclude that most of the absorption will occur close to the jet edges and less at the jet axis. This can be explained physically by a latitudinal density gradient in the jet with a decreasing density towards the jet axis. Hence, the jet has a low-velocity, high-density outflow at the edges and a high-velocity, low-density outflow along the jet axis. 

This jet profile is similar to the configuration discussed in \citet{romanova09}, where the outflow in accreting T Tauri stars consists of a conical-type wind and a magnetically dominated axial jet. The conical wind has velocities close to the Keplerian velocity in the disk from which the wind is ejected, and has a higher density than the axial jet. The magnetically dominated axial jet has a low density and is accelerated towards high velocities. Whether a similar physical process is working in the evolved binary \bd\, is difficult to test as the strength of the magnetic field is not constrained. Moreover, we can show that the jet of \bd\, is not strongly collimated. By applying the assumed mass ratio of $q = 0.6$ in the mass function, we found that the inclination angle of the binary system must be higher than at least $\sim 37^\circ$, such that the mass of the evolved star would be lower than the maximum mass of a white dwarf ($\approx1.4\, M_\odot$). Hence, with this lower limit for inclination and from the relation between the inclination and jet angle, it follows that the jet cannot be strongly collimated. It is more a confined wind with a wide opening angle than a strongly collimated jet. These confined winds have previously been suggested to be present in PPNe and PNe as well \citep[e.g.][]{soker03, soker04}. \citet{soker04} in particular showed that the pairs of fat bubbles observed in several PNe are formed by jets with a half opening angle $\geq\,40^\circ$. Additionally, wide outflows originating from a disk around the companion have been implemented in simulations by \citet{akashi08, akashi13, akashi16} in order to reproduce the shapes of these PPNe and PNe. Hence, these theoretical studies are in good agreement with our observational results.\\

Some features still remain unconstrained, such as the observed absorption at a redshifted RV of $\sim 60\,$km s$^{-1}$ (lower panel of Fig. \ref{fig:v2result}). This absorption might have a physical origin, but could also be due too the subtraction of the template spectrum, creating a non-physical absorption feature.

\section{Doppler tomography}\label{doptom}
The double-peaked emission profile of H$\alpha$ and its phase-dependent variations contain information on the intensity distribution and structure of the circum-companion disk. These orbital variations are studied through the technique of Doppler tomography. A more detailed description of Doppler tomography is provided by \citet{marsh88}. The technique allows us to image the accretion disk indirectly, such that asymmetric structures in the disk can be identified. In order to construct the Doppler map, we used the maximum entropy method that is included in the TRM-DOPPLER\footnote{The TRM-DOPPLER software package can be obtained from https://github.com/trmrsh/trm-doppler.} software package written by Tom Marsh.

Figure \ref{fig:geomdopmap} illustrates the system in spatial coordinates in the co-rotating frame of the binary system. We define phase $\phi =0$ at inferior conjunction, when the primary is in between us and the secondary. This corresponds to the observer being positioned at the positive x-direction. Throughout the phase, the location of the observer rotates clockwise in the co-rotating frame. We construct the Doppler map following this geometry, which is shown at the right in Fig. \ref{fig:geomdopmap}. We note that the velocity curve of the mass stream reaches higher radial velocities than at the origin, which is due to the acceleration of the mass stream from the L1-point towards the accretion disk. The second curve originates from a higher position on the $V_y$-axis and represents the Keplerian velocities that the accretion disk would have at the position of the mass stream. We note that the highest Keplerian velocities are reached in the inner disk and are projected farther out from the origin on the velocity map. Hence, the velocity map represents the accretion disk inside-out.\\

The photospheric corrected H$\alpha$ spectra were used in order to produce the Doppler maps. We mention that these maps are very sensitive to the input data. When we excluded or included different wavelength regions, we noted that the asymmetry in the Doppler map varied significantly. Hence, we examined the considered wavelength ranges and the spectra thoroughly before initiating the calculations of the Doppler maps.
%*** Nakijken *** Hence, the considered wavelength ranges and spectra were thoroughly examined before initiating the calculations of the Doppler maps.

Eleven spectra were excluded because they had a low signal-to-noise ratio, which causes artefacts in the Doppler maps. The excluded spectra are $N=5,11,14,19,21,52,69,79,85,86,$ and 103 of Table \ref{tab:BDradvel}. Telluric lines and cosmic hits are removed in the remaining spectra. The spectra between orbital phase $\phi = 0.25$ and $\phi = 0.75$, which contain the extended blueshifted H$\alpha$ absorption feature, are also excluded, since this absorption feature prevents us from analysing the map properly.

\begin{figure}[t]
\centering
\includegraphics[width =.5\textwidth]{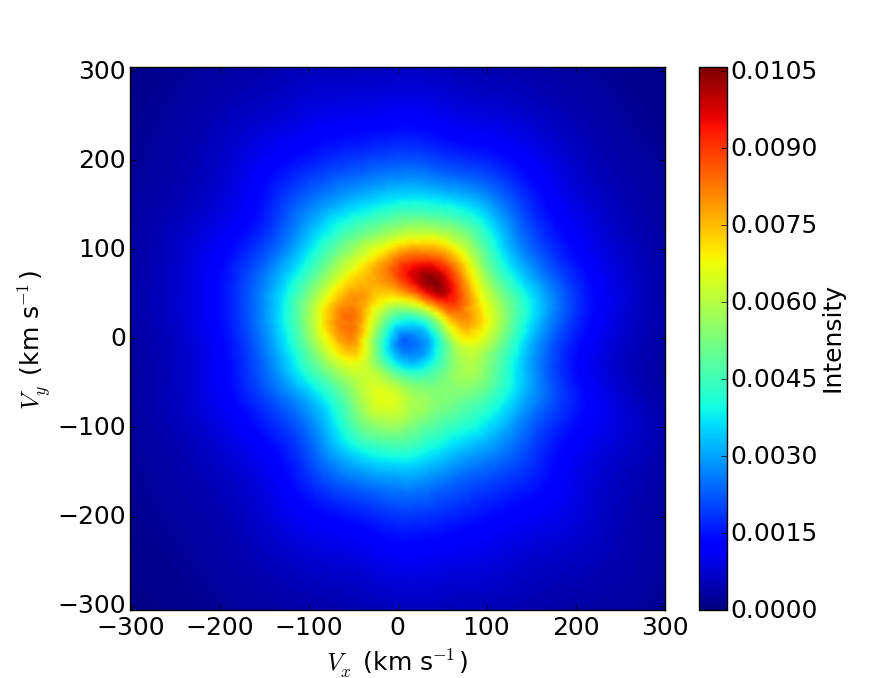}
\caption{Doppler tomograph of \bd\, in H$\alpha$. The colours indicate the flux contribution. }\label{fig:bddopmap}
\end{figure}

\begin{figure*}[t]
\centering
\includegraphics[width =1.0\textwidth]{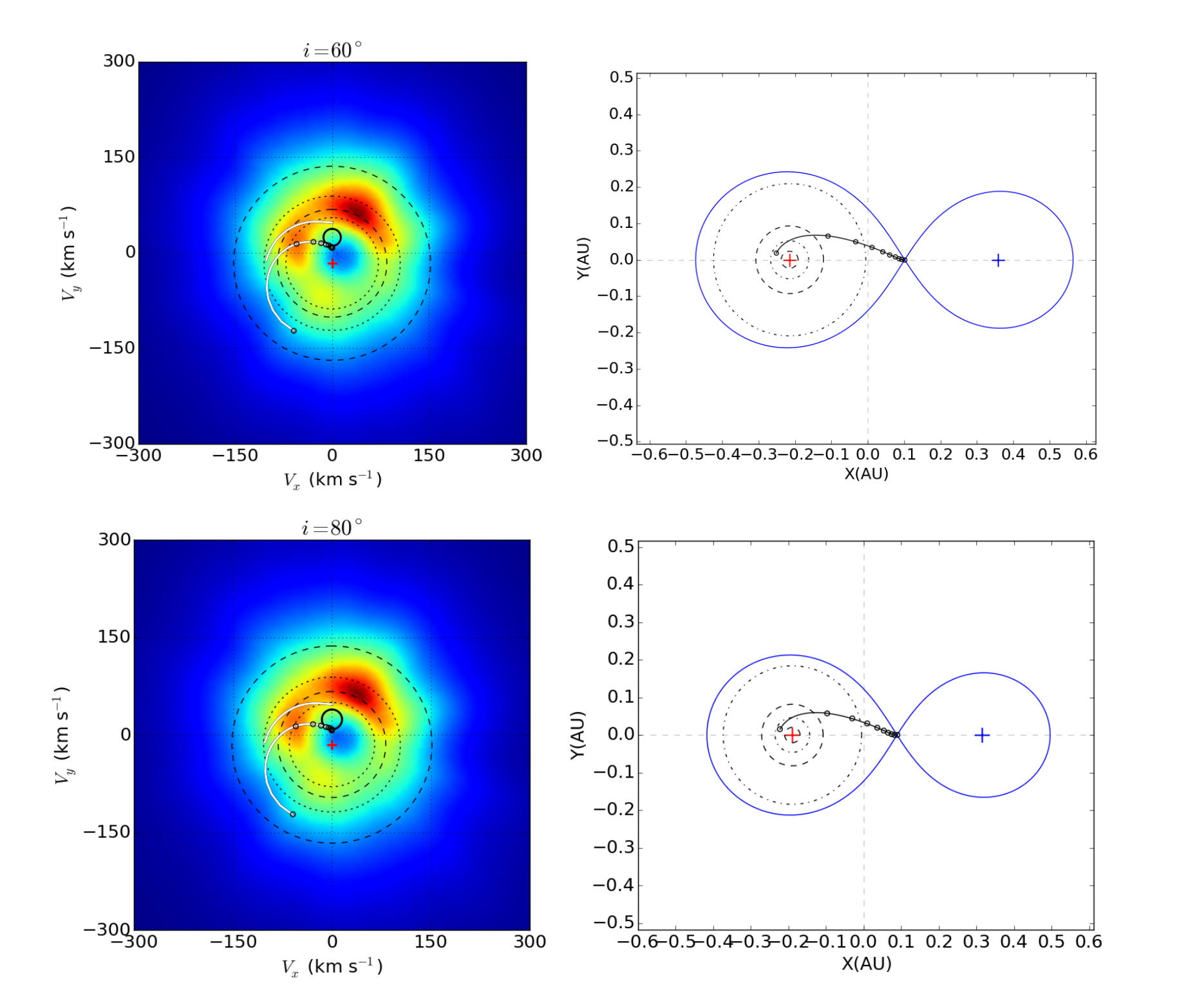}
\caption{Doppler tomograph of \bd\, in H$\alpha$ (left) and the corresponding accretion geometry (right). The map is overplotted with the velocity coordinates of the model for inclination angles of 60$^\circ$ (top left) and 80$^\circ$ (bottom left). The curves shown are similar as for Fig. \ref{fig:geomdopmap}. The points marked on the ballistic trajectory in the geometrical image correspond to the points marked on the curve representing the ballistic trajectory of the gas stream in the Doppler map. The four circles on the Doppler map represent, from outside to inside, a projected Keplerian velocity of 150, 100, 75, and $50\,$km s$^{-1}$, respectively. These correspond to the four circles on the geometrical image.}\label{fig:bddopmapall}
\end{figure*}

By excluding the discussed wavelength regions and the observations with a low signal-to-noise ratio, we improved the quality of our Doppler map. The resulting map is shown in Fig. \ref{fig:bddopmap}.
The map shows a circular structure with an intensity peaking at radial velocities between $40-100\,$km s$^{-1}$. The strongest emission is located in the upper-right quadrant of the map. In order to relate the features in this Doppler map with the structure of the accretion disk, we plot the velocity coordinates of Fig. %\ref{fig:geomvel}
\ref{fig:geomdopmap} over the Doppler map. This is done for two inclination angles of the system and is shown in Fig. \ref{fig:bddopmapall}.
First, we consider the disk velocities. The dashed and dotted curves, which represent the projected Keplerian velocities within the accretion disk, give a good indication at which velocities we observe the strongest emission. The inner disk rim (represented by the outer velocities in the Doppler map) reaches projected velocities of up to $150\,$km s$^{-1}$. The largest flux contribution is located at the upper-right quadrant of the Doppler map, with a second peak at the upper-left quadrant. The location and strength of these peaks can be related to different structures in the accretion disk, such as a hot spot or spiral shocks \citep{marsh01}. A hot spot in the accretion disk will be located at the position where a mass stream hits the accretion disk and causes an increased flux, which can be identified in the Doppler map as an emission peak. This has, for instance, revealed a hot spot in the accretion disk of CE-315 \citep{steeghs04}. The observed bright emission spot in the Doppler map of this cataclysmic variable overlaps the velocity curve of the mass stream, which indicates that the emission peak is related to a hot spot in the disk. 

We note that the second brightest emission peak in the Doppler map of \bd\, overlaps the curves representing the gas stream. This could be an indication of a hot spot in the accretion disk around the secondary component, at a radius where the matter reaches projected Keplerian velocities of 50-75$\,$km s$^{-1}$. The brightest emission peak, located at the upper-right quadrant, does not overlap these curves, however. Hence, a hot spot caused by a mass stream from the L1-point to the accretion disk cannot be the origin of this emission peak. Other sources for this emission peak in the Doppler map could be an asymmetrical structure of the disk itself, or a mass stream from the circumbinary disk onto the secondary component. We wish to emphasize here that since the spectra between orbital phase $\phi = 0.25$ and $\phi = 0.75$ were excluded, the system is only viewed at half the orbital phase. This also influenced the construction of the Doppler map and might cause the resulting emission peaks in the map.\\

Our Doppler images resolve the accretion disk and the velocity structures in it. The second brightest emission peak could correspond to a hot spot at the location where a putative gas stream from the primary would feed the accretion disk. This is not clear, however, since the asymmetrical structure of the Doppler map could also imply that the accretion disk has an asymmetrical structure itself, or that this asymmetry is due to the exclusion of spectra for half the orbital phase.

\section{Discussion and conclusions}\label{sec:concl}
By performing a quantitative analysis of the post-AGB binary \bd , we showed that an active binary interaction in the system leads to the creation of an accretion disk around the secondary component, facilitating a disk-driven outflow or jet. However, the nature of this binary interaction is not yet conclusively clear. The most probable sources to feed the disk around the secondary component are either a mass stream from the dust disk surrounding the binary system, or the accretion of matter from the evolved star through wind RLOF. A mass stream from the evolved star to the secondary often results in hot spot at the location where the mass stream hits the accretion disk. Doppler tomography is an effective technique to identify these hot spots. This has, for instance, determined a hot spot in the accretion disk of dwarf nova U Geminorum \citep{marsh90} and cataclysmic variable WZ Sagittae \citep{spruit98}. The Doppler map of the H$\alpha$ line of \bd\, shows two emission peaks, one of which might be caused by a mass transfer from the primary feeding the disk. This is not conclusive, however, since the emission peak could also be caused by an asymmetrical structure in the disk, or by the exclusion of spectra that cover half an orbital phase.\\
%The results from the Doppler tomography of the H$\alpha$ line infer that a mass transfer from the primary is the most plausible origin for the gas stream feeding the disk. The curve, which indicates the theoretical velocity of the gas stream from the primary, overlaps with the bright spot in the Doppler map for $i = 60^\circ$ and $i = 65^\circ$, as is shown in Fig. \ref{fig:bddopmapall}. This suggests that the gass stream would hit the accretion disk at this position, causing an increased flux.\\

The jet-launching mechanism is most often explained by the interaction of the disk material with a magnetic field \citep{blandford82}. This magnetic field removes angular momentum from the accretion disk and can significantly contribute to the launching and collimation of jets, as was shown by \citet{nordhaus07}. Some PNe harbour high-velocity outflows that are rather collimated, such as the Necklace \citep{corradi11}, Abell 63 \citep{mitchell07}, NGC6778 \citep{guerrero12}, and ETHOS 1 \citep{miszalski11a}. \citet{tocknell14} have derived the magnetic field strength for these four systems, which is in the order of ten gauss for systems where the jet is launched before the common-envelope (CE) phase (the Necklace, Abell 63, and ETHOS 1), or in the order of hundreds up to a few kilogauss when the CE phase predates the jet ejection, as is the case for NGC6778. The collimation of these outflows is thus likely caused by the magnetic field.

For \bd, no magnetic field is known, however. Our modelling shows that the jet is a confined wind with a wide opening angle, and not a strongly collimated jet, as discussed in Sect. \ref{sec:concgeo}. A significant magnetic field causing a strong collimation of the jet is therefore rather unlikely for this system. 

The velocity structure of the jet can be explained best by a latitudinally dependent structure, with the lowest velocities at the jet edges and the highest velocities along the jet axis (Sect. \ref{sec:jetvel}). 
The highest jet velocities rely strongly on the nature of the secondary component, since they must be of the order of the escape velocity near the surface of the accreting object \citep{livio97}. This implies that the jet is launched from the inner regions of the accretion disk, close to the secondary component. Hence, in order to have an estimate of the order of the escape velocity, we must approximate the mass and radius of the secondary. If we assume the evolved component to have a mass of $M_1 = 0.6\,M_\odot$ (which is the mass of a typical post-AGB star), then the mass of the secondary can be derived from the mass function and depends only on the inclination $i$. This results in a secondary mass of $1.48\,M_\odot$, $0.88\,M_\odot$, or $0.71\,M_\odot$, for inclinations of $40^\circ$, $60^\circ$, or $80^\circ$, respectively. The two most plausible candidates for the secondary component are either an MS star or a WD. We first consider the secondary to be an MS star. The radius of the secondary can be approximated from the mass-radius relation of \citet{demircan91} for low-mass terminal-age MS stars:
\begin{eqnarray}
R_2 \cong 2.00\,R_\odot\cdot \left( \frac{M}{M_\odot}\right)^{0.75}.
\end{eqnarray}
The projected escape velocities for the inclinations and stellar masses given above then become $v_\mathrm{MS,esc}\cdot\cos i \approx 940\,\mathrm{km \,s}^{-1}, 390\,\mathrm{km\,s}^{-1}, \mathrm{and\, } 110\, \mathrm{km \,s}^{-1}$. Hence, for inclinations between $i \approx 45-75^\circ$, the projected escape velocity is of the order of the maximum observed jet velocity ($v_0\cdot \cos i = 400\, \mathrm{km \,s}^{-1}$). Next, we determine the projected escape velocity again, but for a WD instead of an MS star as companion. The escape velocity of a WD is in the order of $5000 \, \mathrm{km \,s}^{-1}$. The projected escape velocity is only of the order of the observed jet velocity, provided that the inclination is in a narrow range between 81$^\circ$ and 87$^\circ$ ($v_\mathrm{WD,esc}\cdot\cos i \approx 760-250\,\mathrm{km \,s}^{-1}$). At these inclinations, and given the mass function, the mass of the WD is 0.7$\,M_\odot$, which leads to a mass ratio of $q \approx 0.86$. This is very unlikely given the absence of a clear  orbital motion in the H$\alpha$ line (see Fig.\ref{fig:primrem}). We do not see symbiotic activity either, which may be expected in accreting WDs. We conclude that the secondary component is most likely an MS star and not a WD.\\
Although jet formation is clearly present in this system, the link to the more embedded sources, namely PPNe and PNe, remains to be studied. In addition to the Red Rectangle and \bd, we have accumulated spectra of many more systems, which shows that the jets are very commonly observed \citep{vanwinckel17}. In our future research, we will investigate the whole sample, which will allow us to place constraints on the accretion and jet creation physical processes and their relation to orbital properties. Systems like \bd\, are bright and offer unique possibilities to study accretion processes and jet creation. Very high spatial resolution imaging via interferometry will yield important constraints on accretion processes and the structure and launching physics of jets.

\begin{acknowledgements}
We thank Orsola De Marco for her useful suggestions and comments. DB would also like to thank Steven Bloemen for his help with the Doppler tomography. DK and HvW acknowledge support from the Research Council of the KU Leuven under grant number GOA/2013/012.
DK acknowledges support of the Research Foundation - Flanders under contract G.0B86.13.
Based on observations obtained with the HERMES spectrograph, which is supported by the Research Foundation - Flanders (FWO), Belgium, the Research Council of KU Leuven, Belgium, the Fonds National de la Recherche Scientifique (F.R.S.-FNRS), Belgium, the Royal Observatory of Belgium, the Observatoire de Gen\`eve, Switzerland and the Thueüringer Landessternwarte Tautenburg, Germany. We acknowledge all observers of the HERMES consortium who contributed to obtaining the time series.
\end{acknowledgements}

\bibliographystyle{aa}
\bibliography{allreferences.bib}

\begin{appendix}
\section{Calculations of the path length through the jet}\label{sec:A}

These are the analytical calculations of the path length through the jet, based on the geometrical model of Fig. \ref{fig:geom}. First, we define the coordinates. The origin is at the centre of mass of the system. The orbital plane of the two components is in the xy-plane. The z-axis is perpendicular to the orbital plane and has the same direction as the orbital angular velocity vector. At phase $\phi = 0$, the primary is positioned on the positive y-axis and the secondary on the negative y-axis. The primary and secondary component follow elliptic orbits and their spatial coordinates, $(X_P(\phi),Y_P(\phi),0)$ and $(X_S(\phi),Y_S(\phi),0)$ for the primary and secondary, respectively, at a given phase are determined by the orbital parameters of the system. \\

The jet is considered as a double cone with aperture $2\alpha$, the position of the secondary $(X_S(\phi),Y_S(\phi),0)$ as its apex and the jet-axis parallel to the z-axis. 
The parametric equation of the cone is given by:
\begin{equation}
x(u,\nu) = X_S(\phi) + u \cos(\nu)\tan(\alpha)\hspace{15pt}  0\leq\nu\leq 2 \pi
\end{equation}
\begin{equation}
y(u,\nu) = Y_S(\phi) + u \sin(\nu)\tan(\alpha) \hspace{17pt} 0\leq\nu\leq 2 \pi
\end{equation}
\begin{equation}
z(u,\nu) = u,   
\end{equation}
with parameter $u$ the height along the jet-axis and $\nu$ the azimuth. From the parametric equation, we can determine the implicit Cartesian equation for the cone:
\begin{equation}
z^2(u,\nu) \tan^2(\alpha) =(x(u,\nu)-X_S(\phi))^2 +(y(u,\nu)-Y_S(u,\nu))^2.\label{eq:parcone}
\end{equation}
The line of sight has a direction $\hat{n}=\left(0,\sin (i), \cos (i)\right)$ and is parallel to the yz-plane. The parametric form of the line of sight is given by
\begin{equation}
x(s(\phi)) = x_p(\phi)\label{eq:xs}
\end{equation}
\begin{equation}
y(s(\phi)) = y_p(\phi) + s(\phi) \sin (i)\label{eq:ys}
\end{equation}
\begin{equation}
z(s(\phi)) = z_p(\phi) + s(\phi) \cos (i),\label{eq:zs}
\end{equation}
with $(x_p(\phi),y_p(\phi),z_p(\phi))$ a point on the surface of the primary, which is considered as a disk (we note that $(x_p(\phi),y_p(\phi),z_p(\phi))$ is not necessarily the centre of the primary $(X_P(\phi),Y_P(\phi),0)$).
\begin{figure}[t]
\centering
\includegraphics[width=.5\textwidth]{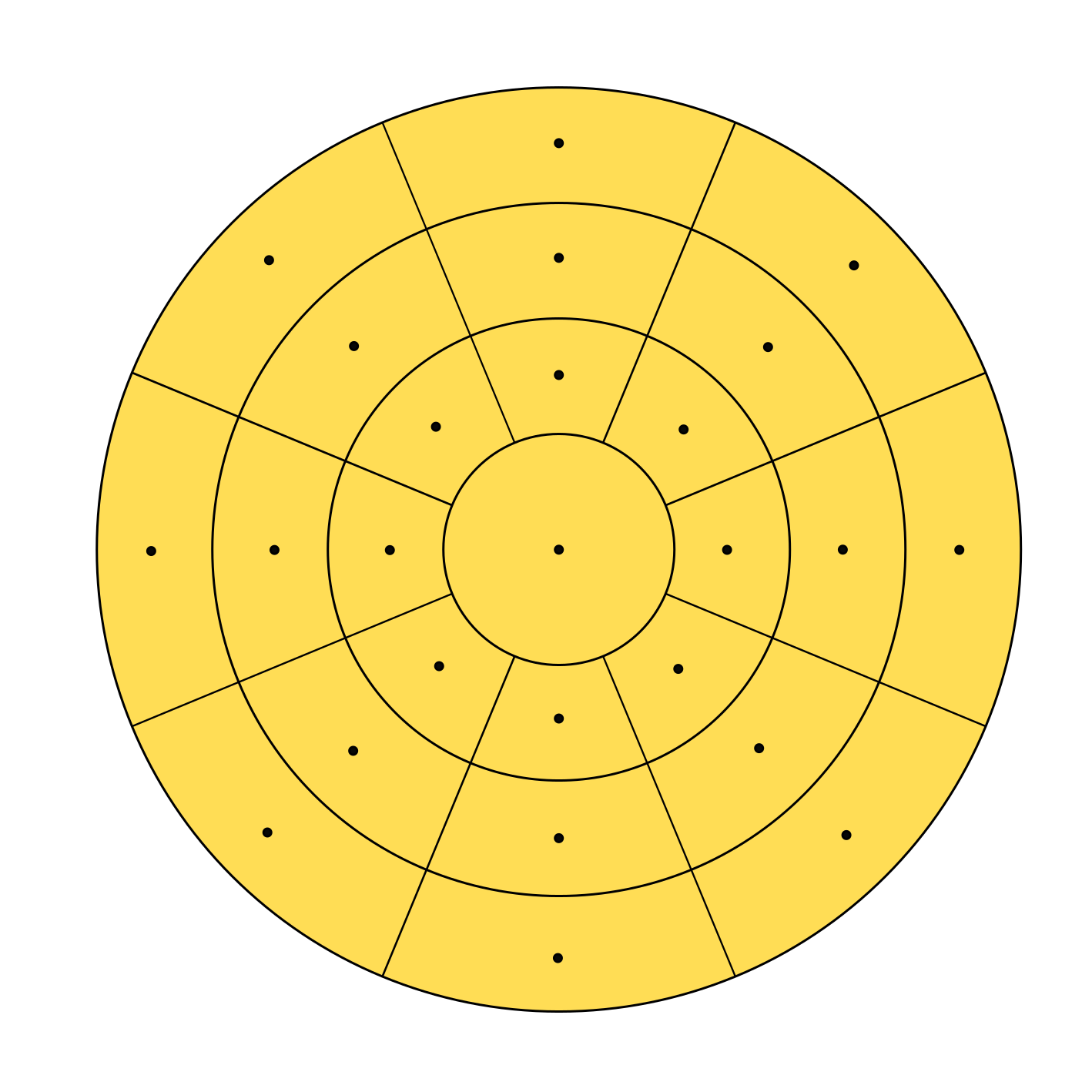}
\caption{Primary represented as a disk in the geometrical model. The disk is divided into 25 parts, with the grid points representing the intensity from each part according to their size.}  \label{fig:disk}
\end{figure}
The path length through the jet is the distance between the two intersection points, $s_1(\phi)$ and $s_2(\phi)$, of the line of sight and the jet. These points can be computed by substituting Eqs. \ref{eq:xs}, \ref{eq:ys} and \ref{eq:zs} in Eq. \ref{eq:parcone}:
\begin{eqnarray}
\left(z_p(\phi)+s(\phi)\cos (i)\right)^2 \tan^2(\alpha)= \left(x_p(\phi)-X_S(\phi)\right)^2\\
+\left(y_p(\phi) + s(\phi) \sin(i)-Y_S(\phi)\right)^2\nonumber
\end{eqnarray}
\begin{eqnarray}
\Leftrightarrow &\left[\tan^2(\alpha)\cos^2(i)-\sin^2(i)\right]\cdot s^2(\phi)\\
&+\left[2z_p(\phi)\tan^2(\alpha)\cos(i)+2(Y_S(\phi)
-y_p(\phi))\sin(i)\right]\cdot  s(\phi)\nonumber\\
&+ z_p^2(\phi)\tan^2(\alpha)-(x_p(\phi) - X_S(\phi))^2-(y_p(\phi)-Y_S(\phi))^2 =0.\nonumber
\end{eqnarray}
Solving this equation for $s(\phi)$ gives the two intersection points:
\begin{eqnarray}
s_1(\phi) = \\
&\frac{-\left[2z_p(\phi)\tan^2(\alpha)\cos(i)+2(Y_S(\phi)-y_p(\phi))\sin(i)\right]+\sqrt{D(\phi)}}{2\left[\tan^2(\alpha)\cos^2(i)-\sin^2(i)\right]}\nonumber
\end{eqnarray}
\begin{eqnarray}
s_2(\phi) = \\
&\frac{-\left[2z_p(\phi)\tan^2(\alpha)\cos(i)+2(Y_S(\phi)-y_p(\phi))\sin(i)\right]-\sqrt{D(\phi)}}{2\left[\tan^2(\alpha)\cos^2(i)-\sin^2(i)\right]}, \nonumber
\end{eqnarray}
with
\begin{eqnarray}
D(\phi) = \\
&4\left[z_p(\phi)\sin(i) \tan(\alpha)-(y_p(\phi)-Y_S(\phi))\tan(\alpha)\cos(i)\right]^2\nonumber \\
&+ 4(x_p(\phi) X_S(\phi))^2 \cdot\left(\tan^2(\alpha)\cos^2(i)-\sin^2(i)\right).\nonumber
\end{eqnarray}
Hence, the path length at each orbital phase is given by $s(\phi) = s_2(\phi)-s_1(\phi)$. \\

The primary, which is an evolved giant, cannot be considered as a point mass because of its relatively large stellar radius. For this reason, we consider the primary in our model as a uniform disk with the same radius as the stellar radius. The uniform disk contains 25 grid points, as is shown in Fig. \ref{fig:disk}. Each grid point represents the surface it is positioned in. The intensity $I_i$ of each part, relative to the total intensity $I_\mathrm{tot}$ of the surface is: 
\begin{equation}
I_i=\frac{A_i}{A_\mathrm{tot}}I_\mathrm{tot},
\end{equation}
with $A_i$ the surface of part $i$ and $A_\mathrm{tot}$ the total surface. Hence, we do not take limb darkening into account. At each orbital phase, the path length is calculated for the line of sight going through each grid point of the disk. Since the grid points represent the intensity from each part of the disk, the path length of all grid points are summed according to their relative weight:
\begin{equation}
s_\mathrm{tot} = \sum_i^{25}\frac{A_i}{A_\mathrm{tot}} s_i. 
\end{equation}
\section{Radial velocities}
\longtab{1}{
\begin{longtable}{ccccc}
\caption{Exposure time and radial velocities of BD+46$\degr$442.}\label{tab:BDradvel} \\
\hline
\hline 
$N$ & BJD & Exp. & RV & $\sigma$ \\
 & $-2455000\,$d & s & km\,s$^{-1}$ &  km\,s$^{-1}$ \\
 \hline
\endfirsthead
\caption*{\textbf{Table \ref{tab:BDradvel}} Continued}\\
\hline
\hline 
$N$ & BJD & Exp. & RV & $\sigma$ \\
 & $-2455000\,$d & s & km\,s$^{-1}$ &  km\,s$^{-1}$ \\
 \hline
\endhead
\hline
\endfoot
1 & 31.7079 & 1800 & -88.02  & 0.23 \\
2 & 42.6956 & 1200 & -78.38  & 0.26 \\
3 & 60.6997   & 1800 & -74.86  & 0.22   \\
4 & 60.7211 & 1800 & -73.11  & 0.37 \\
5 & 85.6417 & 530 & -91.14  & 0.22 \\
6 & 92.6438 & 520 & -97.11  & 0.24 \\
7 & 100.6184 & 522 & -108.44 & 0.16 \\
8 & 125.4916 & 525 & -122.69 & 0.22 \\
9 & 131.5309 & 800 & -120.16 & 0.26 \\
10 & 159.5515 & 800 & -99.10  & 0.56 \\
11 & 159.5611 & 653 & -99.57  & 0.56 \\
12 & 199.3349 & 1000 & -73.34  & 0.45 \\
13 & 216.3841 & 900 & -81.22  & 0.37 \\
14 & 224.3558 & 900 & -88.57  & 0.30 \\
15 & 421.6270 & 700 & -116.60 & 0.22 \\
16 & 425.6729 & 600 & -112.47 & 0.26 \\
17 & 432.5857 & 600 & -106.70 & 0.21 \\
18 & 468.6638 & 721 & -75.74  & 0.31 \\
19 & 471.6246 & 600 & -74.63  & 0.57 \\
20 & 499.5413 & 600 & -81.54  & 0.14 \\
21 & 507.4677 & 1200 & -88.58  & 0.21 \\
22 & 553.4767 & 900 & -121.98 & 0.14 \\
23 & 573.3428 & 900 & -106.24 & 0.26 \\
24 & 581.3361 & 1000 & -99.42  & 0.19 \\
25 & 767.7082 & 900 & -74.87  & 0.28 \\
26 & 769.7095 & 900 & -75.62  & 0.26 \\
27 & 776.7314 & 900 & -79.69  & 0.43 \\
28 & 783.6924 & 900 & -84.01  & 0.37 \\
29 & 791.7376   & 900 & -94.33  & 0.10      \\
30 & 797.7256 & 900 & -99.14  & 0.09 \\
31 & 826.6507 & 1150 & -122.48 & 0.23 \\
32 & 835.5716 & 900 & -120.55 & 0.12     \\
33 & 841.4968 & 1200 & -118.46 & 0.17 \\
34 & 842.5869 & 900 & -117.06 & 0.20 \\
35 & 845.6115 & 900 & -114.07 & 0.24 \\
36 & 848.4110 & 900 & -112.27 & 0.18 \\
37 & 852.5181 & 1200 & -109.07 & 0.20 \\
38 & 857.6440 & 900 & -103.89 & 0.17 \\
39 & 860.5961 & 900 & -101.69 & 0.20 \\
40 & 866.5383 & 900 & -96.54  & 0.18 \\
41 & 870.5593 & 900 & -93.90  & 0.34 \\
42 & 872.6056 & 900 & -91.35  & 0.19 \\
43 & 880.5575 & 900 & -82.86  & 0.15 \\
44 & 882.4635 & 900 & -82.28  & 0.28 \\
45 & 884.5534 & 900 & -81.31  & 0.32 \\
46 & 886.4946 & 900 & -80.58  & 0.24 \\
47 & 888.5519 & 1027 & -80.09  & 0.25 \\
48 & 890.5195 & 1300 & -79.45  & 0.22 \\
49 & 890.5351 & 1300 & -79.70  & 0.20 \\
50 & 903.3808 & 900 & -74.18  & 0.17 \\
51 & 911.4050 & 1600 & -75.50  & 0.34 \\
52 & 930.3926  & 1200 & (...)  &  (...)\\
53 & 934.4143  & 900 & -96.28  & 0.14 \\
54 & 940.3635 & 1300 & -103.25 & 0.39 \\
55 & 943.4399 & 900 & -107.07 & 0.27 \\
56 & 948.3964 & 900 & -112.70 & 0.37 \\
57 & 953.3599   & 900 & -117.99 & 0.15      \\
58 & 956.3969  & 900 & -118.22 & 0.37 \\
59 & 957.4816 & 1000 & -118.98 & 0.39 \\
60 & 968.4543 & 750 & -120.74 & 0.49 \\
61 & 1127.6495 & 700 & -114.17 & 0.46 \\
62 & 1134.6397 & 700 & -107.97 & 0.37 \\
63 & 1152.6995 & 660 & -93.23  & 0.13 \\
64 & 1166.6584 & 900 & -82.07  & 0.23 \\
65 & 1179.5528 & 940 & -74.35  & 0.13 \\
66 & 1194.6011 & 900 & -75.15  & 0.19 \\
67 & 1199.6301 & 900 & -77.95  & 0.24 \\
68 & 1200.6435 & 900 & -78.75  & 0.22 \\
69 & 1212.5012 & 900 & -90.66  & 0.22 \\
70 & 1247.4768 & 900 & -122.67 & 0.20 \\
71 & 1266.4321 & 900 & -114.01 & 0.29 \\
72 & 1306.3493 & 900 & -82.31  & 0.36 \\
73 & 1320.3713 & 900 & -75.09  & 0.25 \\
74 & 1322.3294 & 900 & -74.35  & 0.17 \\
75 & 1324.3398 & 1200 & -74.13  & 0.22 \\
76 & 1332.3630 & 1000 & -74.08  & 0.15 \\
77 & 1470.7074 & 900 & -73.56  & 0.25 \\
78 & 1484.6901 & 950 & -81.28  & 0.17 \\
79 & 1505.7255 & 900 & -105.43 & 0.22 \\
80 & 1515.7182 & 660 & -114.61 & 0.18 \\
81 & 1531.6971 & 900 & -122.19 & 0.15 \\
82 & 1559.6417 & 900 & -103.66 & 0.14 \\
83 & 1565.7152 & 900 & -99.82  & 0.13 \\
84 & 1597.6389 & 900 & -74.19  & 0.18 \\
85 & 1611.5700 & 900 & -73.13  & 0.31 \\
86 & 1631.5401 & 900 & -87.94  & 0.19 \\
87 & 1661.3673 & 900 & -118.27 & 0.32 \\
88 & 1670.3677 & 900 & -121.08 & 0.27 \\
89 & 1700.3739 & 900 & -105.30 & 0.11 \\
90 & 1869.7322 & 900 & -82.32  & 0.13 \\
91 & 1894.6891 & 900 & -73.47  & 0.28 \\
92 & 1905.6457  & 810 & -80.74  & 0.10      \\
93 & 1908.5946 & 600 & -82.98  & 0.17 \\
94 & 1930.6295 & 900 & -105.96 & 0.17 \\
95 & 1961.4932 & 900 & -122.17 & 0.20 \\
96 & 1995.3929 & 900 & -93.90  & 0.11 \\
97 & 2027.3705 & 700 & -72.85  & 0.18 \\
98 & 2057.4416 & 700 & -91.29  & 0.18 \\
99 & 2208.7132 & 900 & -104.23 & 0.25 \\
100 & 2238.6683 & 900 & -120.44 & 0.16 \\
101 & 2267.6976 & 680 & -102.67 & 0.14 \\
102 & 2268.6904 & 900 & -101.90 & 0.13 \\
103 & 2342.5220 & 750 & -94.39  & 0.19 \\
104 & 2400.3956 & 1800 & -108.86 & 0.21
\end{longtable}
}
\end{appendix}
\end{document}